\documentclass[prd,preprint,nofootinbib]{revtex4-1}

\usepackage{color}
\usepackage[latin1]{inputenc}
\usepackage[T1]{fontenc}
\usepackage{amsmath}
\usepackage{graphicx}
\usepackage{epstopdf}
\graphicspath{{figures//}}

\begin{document}

\title{Debye screening mass near deconfinement from holography}

\author{S.~I.~Finazzo}
\email{stefano@if.usp.br}
\affiliation{Instituto de F\'{i}sica, Universidade de S\~{a}o Paulo, S\~{a}o Paulo, SP, Brazil}

\author{J.~Noronha}
\email{noronha@if.usp.br}
\affiliation{Instituto de F\'{i}sica, Universidade de S\~{a}o Paulo, S\~{a}o Paulo, SP, Brazil}
\date{\today}

\begin{abstract}
In this paper the smallest thermal screening mass associated with the correlator of the $CT$-odd operator, $\sim {\rm Tr}F_{\mu\nu}\tilde{F}^{\mu\nu}$, is determined in strongly coupled non-Abelian gauge plasmas which are holographically dual to non-conformal, bottom-up Einstein+scalar gravity theories. These holographic models are constructed to describe the thermodynamical properties of $SU(N_c)$ plasmas near deconfinement at large $N_c$ and we identify this thermal mass with the Debye screening mass $m_D$. In this class of non-conformal models with a first order deconfinement transition at $T_c$, $m_D/T$ displays the same behavior found for the expectation value of the Polyakov loop (which we also compute) jumping from zero below $T_c$ to a nonzero value just above the transition. In the case of a crossover phase transition, $m_D/T$ has a minimum similar to that found for the speed of sound squared $c_s^2$. This holographic framework is also used to evaluate $m_D$ as a function of $\eta/s$ in a strongly coupled conformal gauge plasma dual to Gauss-Bonnet gravity. In this case, $m_D/T$ decreases with increasing $\eta/s$ in accordance with extrapolations from weak coupling calculations.
\end{abstract}

\maketitle

\tableofcontents

\section{Introduction}

In the deconfined phase of non-Abelian gauge theories, the inverse of the Debye screening mass, $m_D^{-1}$, can be used to define a screening length of the thermal medium that roughly signals the effective maximum interaction distance between two colored heavy probes. Debye screening is the mechanism behind Matsui and Satz's well known proposal \cite{Matsui:1986dk} that the ``melting" (dissociation) of heavy quarkonia states in the QGP is a signature of deconfinement. 

Although in weakly coupled Abelian and non-Abelian plasmas the Debye screening mass has been calculated long ago at one loop in perturbation theory \cite{Shuryak:1980tp,Gross:1980br,Kapusta:2006pm}, higher order perturbative calculations \cite{Nadkarni:1986cz,Rebhan:1993az,Braaten:1994pk,Rebhan:1994mx} indicate the breakdown of the perturbation series expansion for this quantity. Thus, a non-perturbative, gauge invariant definition of the Debye screening mass is needed. A definition that is inherently non-perturbative and gauge invariant was proposed by Arnold and Yaffe in Ref.\ \cite{Arnold:1995bh} where $m_D$ was defined as the largest inverse screening length among all the possible Euclidean correlation functions involving pairs of $CT$-odd operators in the thermal gauge field theory. Previous studies concerning thermal screening lengths in non-Abelian plasmas include lattice calculations \cite{Kajantie:1997pd,Datta:1999yu,Datta:1998eb,Datta:2002je,Laine:1999hh,Hart:2000ha,Nakamura:2003pu}, non-perturbative analyses of the gluon propagator at finite temperature \cite{Cucchieri:2000cy,Cucchieri:2012gb,Aouane:2012bk,Silva:2013maa}, other analytical approaches \cite{Laine:2009dh,Chakraborty:2011uw}, and holographic calculations \cite{Bak:2007fk,Hoyos:2011uh,Singh:2012xj}.

In this paper we use the gauge/gravity duality \cite{Maldacena:1997re,Witten:1998qj,Witten:1998zw,Gubser:1998bc} to understand the general properties of the smallest thermal screening mass associated with the $CT$-odd operator, $\sim {\rm Tr}F_{\mu\nu}\tilde{F}^{\mu\nu}$, in non-conformal strongly coupled plasmas described by Einstein gravity plus a scalar field. We shall follow \cite{Bak:2007fk} and {\it identify} this thermal screening mass as the Debye mass $m_D$ in the strongly-coupled plasma. After associating this Debye screening mass in the field theory with the lowest lying mass in the spectrum \cite{Csaki:1998qr,deMelloKoch:1998qs} of axion fluctuations in the bulk \cite{Bak:2007fk}, we show (given some reasonable conditions regarding the axion effective action) that the bulk axion spectrum is gapped, positive, and discrete in the deconfined phase of these theories. This shows that this thermal screening radius, which may be relevant for the melting of heavy quarkonia in this class of strongly-coupled large $N_c$ plasmas, is necessarily finite (even in the case of a second order deconfining transition). Also, we find that $m_D/T$ generally follows the behavior of the expectation value of the Polyakov loop operator near the phase transition. In fact, for a first order deconfinement phase transition $m_D/T$ jumps from zero below the critical temperature $T_c$ to a finite value immediately above it. 

To estimate the behavior of this screening mass in a non-conformal strongly coupled plasma with similar properties to the QCD plasma, we consider a variety of holographic bottom-up models constructed using 5 dimensional Einstein + scalar effective bulk actions. The first model, which we call Model A, is built in the context of Improved Holographic QCD (IHQCD) \cite{Gursoy:2007cb,Gursoy:2007er,Gursoy:2008bu,Gursoy:2008za,Gursoy:2009jd}, being a simple analytical model \cite{Kajantie:2011nx} involving an Einstein+scalar gravity bulk action dual to a strongly coupled non-Abelian which possesses a first order confinement/deconfinement phase transition. The second class of models (Model B) \cite{Gubser:2008ny,Gubser:2008yx,Gubser:2008sz} are also based on Einstein+scalar bulk actions but now the scalar potentials are chosen in order to reproduce some lattice QCD thermodynamical results. The model that reproduces lattice data for pure SU(3) Yang-Mills, which possesses a first order deconfinement transition \cite{Boyd:1996bx,Panero:2009tv,Borsanyi:2012ve}, is called Model B1, whereas the model that matches lattice data for QCD with (2+1) light flavors of quarks \cite{Borsanyi:2010cj} is called Model B2. For all models, A, B1, and B2, we obtain, numerically, the screening mass $m_D$ as a function of the temperature $T$. For models A and B1, both of which present a first order phase transition, we explicitly verify the existence of a discontinuity in $m_D/T$ at the critical temperature $T_c$, where $m_D/T$ jumps discontinuously from 0 to a finite value above $T_c$. For the model B2, which displays a crossover phase transition, $m_D/T$ increases with $T$ smoothly from 0 and has a local minimum at a given temperature (following a behavior similar to that shown by the speed of sound), after which it then continuously rises to its conformal limit.

As a final application, we consider the screening mass in a strongly coupled conformal plasma dual to Gauss-Bonnet gravity \cite{Zwiebach:1985uq,Cai:2001dz}. In this theory the shear viscosity to entropy density ratio, $\eta/s$, is different than $1/(4\pi)$ \cite{Policastro:2001yc,Buchel:2003tz,Kovtun:2004de} for a range of values of the controlling parameter of the theory, $\lambda_{GB}$, associated with the higher order derivatives in the action as shown in \cite{Brigante:2008gz,Brigante:2007nu}. Thus, in this case one can see how $m_D/T$ depends upon $\eta/s$ in this strongly coupled plasma and compare with the results of the phenomenological approach based on fits to the heavy quark potential at strong coupling pursued in Ref. \cite{Finazzo:2013rqy}. We find the intriguing result that $m_D/T$ decreases with increasing $\eta/s$.

This paper is organized as follows. In Section \ref{sec:gen} we motivate the non-perturbative definition of thermal screening lengths in non-Abelian gauge theories (the reader that is already familiar with Ref.\ \cite{Arnold:1995bh} may want to skip the introductory sections \ref{IIa} and \ref{IIb} and go directly to \ref{IIc}) and present the holographic prescription for evaluating these quantities in strongly coupled plasmas dual to bottom-up theories of gravity involving the metric and a scalar field. In this section we also present some general results for the thermal screening mass associated with ${\rm Tr}F_{\mu\nu}\tilde{F}^{\mu\nu}$ which are valid in this holographic framework. In Section \ref{sec:adsdebye} we briefly review the results and techniques of Refs. \cite{Bak:2007fk,Csaki:1998qr,deMelloKoch:1998qs} for evaluating this thermal screening mass in a strongly coupled $\mathcal{N}=4$ SYM plasma. Section \ref{sec:ihqcd} is dedicated to the evaluation of $m_D$ and the Polyakov loop in Model A. In Section \ref{ref:Bclass} we review some general results for the B class of models pertinent to our purposes. Section \ref{sec:B1model} (Section \ref{sec:B2model}) is reserved for the evaluation of $m_D$ in the B1 Model (B2 Model, respectively). We show that the heavy quark free energy (extracted from the expectation value of the Polyakov loop) in these holographic models for $SU(N_c)$ Yang-Mills theory nicely describes recent lattice data \cite{Mykkanen:2012ri}. In Section \ref{sec:gaussbonnet} we analyze $m_D \times \eta/s$ in Gauss-Bonnet gravity. Section \ref{sec:conclusions} contains our conclusions and outlook\footnote{In Appendix \ref{sec:appendix-gauge} we present the technical details of a coordinate change used in the study of the B models. We also present the evaluation of the glueball spectrum at $T=0$ in Model B1 in Appendix \ref{sec:appendix-spectra}.}.

\section{General Results For The Holographic Debye Screening Mass}
\label{sec:gen}

For the sake of completeness, in Sections \ref{IIa} and \ref{IIb} we review some necessary results on screening lengths in thermal gauge theories and the non-perturbative definition of the Debye screening mass proposed in \cite{Arnold:1995bh}. Then, in \ref{IIc} we motivate the holographic prescription for the evaluation of the Debye mass and study some of its general properties using holography.

\subsection{Screening lengths in thermal gauge theories}\label{IIa}

Let $\hat{O}$ be a gauge invariant operator and consider the (equal-time) Euclidean 2-point correlation function
\begin{equation}
\label{eq:correlator}
G_E (\vec{x}) \equiv \langle 0 | \hat{O}^{\dagger} (\vec{x}) \hat{O}(\vec{0}) | 0 \rangle.
\end{equation}
A QFT in thermal equilibrium can, as usual, be studied using the Matsubara (or imaginary time) formalism \cite{Kapusta:2006pm}, where we consider the compactification of the imaginary time $\tau = i t$ direction in a circle of radius $\beta = 1/T$, where $T$ is the temperature of the thermal bath. A key insight to this discussion \cite{Arnold:1995bh,Bak:2007fk} is that the resulting Euclidean symmetry allows us, instead of compactifying along the time direction, to compactify along any of the spatial directions; for instance, we may compactify along the $x$ spatial direction. Let $\left\{|n\rangle \right\}$ be a complete set of eigenstates of the translation operator $H_E$ along the $x$ direction, with corresponding eigenvalues $E_n$. Then, inserting the completeness relation for the basis $\left\{|n\rangle \right\}$ one finds
\begin{equation}
\label{eq:correlator2}
G_E ( x) = \sum\limits_{n=0}^{\infty} \langle 0 |\hat{O}^{\dagger} (x)| n \rangle \langle n|\hat{O}(0)  | 0 \rangle.
\end{equation}
Since $H_E$ is an Euclidean translation operator
\begin{equation}
\label{eq:trans}
\hat{O} (x) = e^{H_E |x|} \hat{O}(0) e^{-H_E |x|}
\end{equation}
and, thus,
\begin{equation}
\label{eq:correlator3}
G_E (x) = \sum\limits_{n=0}^{\infty} e^{-E_n |x|} |c_n|^2,
\end{equation}
where
\begin{equation}
\label{eq:coef}
c_n \equiv \langle n | \hat{O} (0) |0 \rangle.
\end{equation}
For large spatial separations, the ground state contribution to Eq.\ \eqref{eq:correlator3} dominates and
\begin{equation}
\label{eq:correlator4}
G_E (x) \sim e^{-E_0 |x|} |c_0|^2.
\end{equation}
Thus, $E_0^{-1}$ may be taken as the screening length of $G_E(x)$ - for distances $|x|$ greater than $E_0^{-1}$ the fluctuations of $\hat{O}$ are effectively not correlated.

\subsection{Non-perturbative definition of the Debye screening mass}\label{IIb}

In this section we briefly review the non-perturbative definition of the Debye screening mass proposed in \cite{Arnold:1995bh}.

\begin{figure}
\centering
  \includegraphics[width=.3\linewidth]{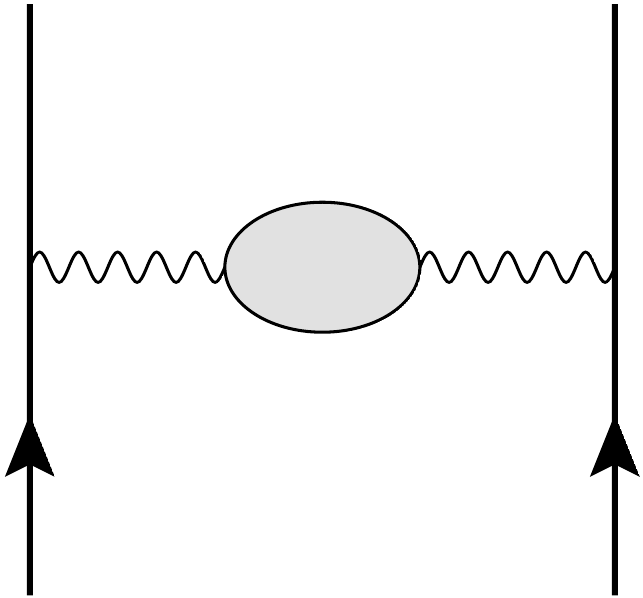}
  \caption{Perturbative definition of the Debye mass. A single photon (gluon) is exchanged between two static test charges. The pole of the photon (gluon) propagator at zero frequency gives the Debye screening mass $m_D$, the inverse screening length of the static potential.}
  \label{fig:debyemass}      
\end{figure}

In quantum electrodynamics (QED), perturbatively, the Debye screening mass $m_D$ can be determined as the pole in the $00$ component of the photon propagator at zero frequency, $\Pi_{00} (0,\vec{p}^{\,2})$ (Fig. \ref{fig:debyemass}) - i.e., the solution of
\begin{equation}
\label{eq:debyemassdef}
\Pi_{00} (0,\vec{p}^{\,2}=-m_D^2) + m_D^2 = 0.
\end{equation}
The screening length of the static potential of two static test charges is given by the inverse Debye mass $m_D^{-1}$. Magnetic fields are unscreened in perturbation theory so that $\Pi_{ij} \to 0$ as $\vec{p} \to 0$ - the Debye screening mass captures the physics of electric screening. This definition can be applied perturbatively to non-Abelian gauge theories, yielding the lowest order, one-loop, perturbative result in the ultrarelativistic approximation (neglecting particle masses and chemical potentials) \cite{Shuryak:1980tp,Gross:1980br,Kapusta:2006pm}
\begin{equation}
\label{eq:pertmD}
m_D = \sqrt{\frac{N_c}{3}+\frac{N_f}{3}} g T + O(g^2T),
\end{equation}
for an $SU(N_c)$ gauge theory with $N_f$ minimally coupled fermions, where $g$ is the gauge theory coupling constant.

Ref.\ \cite{Arnold:1995bh} proposed a way to define the Debye screening mass in an explicit gauge invariant (and non-perturbative) manner using Euclidean time reflection symmetry that is useful in the context of strongly-coupled plasmas. Consider the $CT$ (the composite of time reversal $T$ and charge conjugation $C$) transformation in real time. The corresponding symmetry in Euclidean time is $R_{\tau}$, where $R_{\tau}$ is the imaginary (Euclidean) time reflection. To see this, note that any Lorentz invariant theory must have $CPT$ symmetry, where $P$ stands for spatial inversion. Correspondingly, any Euclidean invariant theory must be rotation invariant. Since $P R_{\tau}$ is a pure rotation in an Euclidean theory, $CPT$ must correspond to $P R_{\tau}$. Also given that $P$ is time independent, $R_{\tau}$ must correspond to $CT$. Since $A_0$ is even under $R_{\tau}$ and $\vec{A}$ is odd under $R_{\tau}$, the authors of Ref. \cite{Arnold:1995bh} defined the Debye screening mass $m_D$ as the inverse of the largest correlation length (or, equivalently, the smallest screening mass) of all correlation functions $\langle \hat{A} (\vec{x}) \hat{B}(\vec{0}) \rangle$ involving two local, gauge invariant operators $\hat{A}$, $\hat{B}$, both odd under Euclidean time reflection $R_{\tau}$ ($CT$ in real time). This construction explicitly removes the magnetic gluon exchange and takes into account only the chromo-electric gluons. Thus, according to \cite{Arnold:1995bh}, the Debye screening mass may be defined as the largest inverse screening length in this channel
\begin{equation}
\label{eq:debyedef}
G_E(\vec{x}) = \langle \hat{A} (\vec{x}) \hat{B} (\vec{0}) \rangle \sim e^{-m_D |\vec{x}|} \quad \mathrm{as} \,\, |\vec{x}| \to \infty\,.
\end{equation}

In this paper we will adopt this definition of the Debye screening mass since it can be readily used in the case of strongly-coupled plasmas that are holographically dual to theories of gravity, as shown in \cite{Bak:2007fk}. From the preceding discussion, we see that to evaluate this Debye screening mass one has to determine correlation lengths of two point functions in a non-Abelian plasma - or, equivalently, evaluate the smallest $R_{\tau}$ odd glueball mass in a 3 dimensional Yang-Mills theory at zero temperature. From the holographic standpoint, the extraction of the glueball masses in the large $N_c$ and strong coupling limit was done in Refs. \cite{Csaki:1998qr, deMelloKoch:1998qs}. The holographic prescription for evaluating the glueball masses corresponds to analyze in the theory of gravity dual to the QFT in question the fluctuations of the bulk fields that source, in the corresponding gauge theory, the gauge invariant operators that couple to the glueballs which have the same quantum numbers of the dual bulk field.

In the case of the $J^{PC} = 0^{-+}$ channel (which is $R_{\tau}$ odd), according to the IHQCD framework \cite{Gursoy:2007cb,Gursoy:2007er,Gursoy:2012bt}, one must analyze the dimension 4 operator $\mathrm{Tr} \, F_{\mu \nu} \tilde{F}^{\mu \nu}$ which is sourced by a massless (pseudoscalar) axion field $a(x,z)$ in the bulk. Then, as discussed in \cite{Bak:2007fk}, the Debye mass corresponds to the imaginary wavevector of smallest magnitude for which the equations of motion corresponding to the axion fluctuations admit plane wave solutions of the kind $e^{i \vec{k}\cdot \vec{x}}\, a(z)$ that are regular at the horizon and obey a Dirichlet condition at the boundary \cite{Kovtun:2005ev}.

A direct consequence of this definition of the Debye screening mass in holographic strongly-coupled plasmas is that this $m_D$ is independent of the gauge coupling and the number of colors when both of them are sufficiently large. In fact, since this $m_D$ is determined by the bulk fluctuations of the axion in a supergravity-like action, this quantity cannot depend on the gauge coupling (since for a two derivative action there are no terms including the string scale $\ell_s$) or the number of colors (which only appears in this case as an overall multiplicative factor in the action in the form of the 5-dimensional Newton's constant). This should be kept in mind when one tries to make a connection between these strongly coupled results and the general intuition acquired over the years about the Debye mass computed within perturbation theory. For instance, we shall show below that this $m_D$ is never zero in the deconfined phase of the strongly-coupled plasma, which is described by a black brane in the bulk. Therefore, even in the case of a second-order phase transition the $m_D$ we compute would be nonzero. Thus, one cannot directly identify this quantity with the one that describes the fluctuations of Polyakov loops in effective models for the quark-gluon plasma \cite{Pisarski:2000eq,Dumitru:2003hp,Pisarski:2006hz,Dumitru:2012fw}.

\subsection{General properties of the holographic axion spectrum}\label{IIc}

Armed with the holographic prescription for extracting the Debye screening mass by means of the bulk axion spectrum, we now examine some of its general properties in a large class of gravity duals. The action for the fluctuations of the massless axion in these backgrounds is assumed to be of the form\footnote{Note that since in IHQCD the bulk axion is trivial in the background, the action for its fluctuations is easily determined to be the one in (\ref{eq:axionaction}) \cite{Gursoy:2012bt}.}
\begin{equation}
\label{eq:axionaction}
S = \frac{1}{32 \pi G_5} \int d^5x \, \sqrt{g}\,\left(\mathcal{Z}(z)\,  g^{\mu \nu} \partial_{\mu} a\, \partial_{\nu} a\right)\,,
\end{equation}
where $G_5 \sim 1/N_c^2$ is the 5-dimensional gravitational constant and $\mathcal{Z}(z)$ is a given function of the holographic coordinate $z$ - the reason for including this axion coupling function is that in certain classes of backgrounds, as in those of Improved Holographic QCD \cite{Gursoy:2007cb,Gursoy:2007er,Gursoy:2008bu,Gursoy:2008za,Gursoy:2009jd}, a resummation of the contributions originating from string theory should result in an effective action for the axion that involves this multiplicative factor. The specific form for this function will be defined later in the paper.

The background metric for the asymptotically AdS$_5$ spacetime (with conformal boundary at $z\to 0$) is defined by the line element
\begin{equation}
ds^2  = e^{2\mathcal{A}(z)}\left(f(z)d\tau^2+d\vec{x}^2+\frac{dz^2}{f(z)}\right)
\label{eq:bgmetric}
\end{equation}
where $f(0)=1$ and the black brane horizon $z_h$ is the smallest root of $f(z_h)=0$. Moreover, note that $\lim_{z\to 0}e^{2\mathcal{A}(z)}=R^2/z^2$ where $R$ is the radius of the asymptotic AdS$_5$ space. The equation of motion for the axion is
\begin{equation}
\label{eq:axioneom}
\partial_{\mu} (\mathcal{Z}(z) e^{5\mathcal{A}} g^{\mu \nu} \partial_{\nu} a) = 0
\end{equation}
and, in momentum space (taking the Matsubara frequency to zero since we want the largest inverse correlation length) with the plane wave Ansatz $a(\vec{x},z) \to e^{i\vec{k} \cdot \vec{x}} a(z)$, one finds the equation of motion (with $M^2 = - \vec{k}^2$)
\begin{equation}
\label{eq:axioneom}
\partial_z (e^{2\mathcal{B}} f(z) a') + M^2 e^{2\mathcal{B}} a = 0,
\end{equation}
where $a'(z) = da(z)/dz$ and we have defined the function
\begin{equation}
\label{eq:Bdef}
\mathcal{B}(z) \equiv \frac{3}{2} \mathcal{A}(z) + \frac{1}{2} \log \mathcal{Z}(z) \,.
\end{equation}

An alternative, but useful, form of the equation of motion is obtained by introducing $\psi = e^{\mathcal{B}} a$, which leads to
\begin{equation}
\label{eq:axionsch}
-\psi'' - \frac{f'}{f} \psi' + \frac{1}{f}\left[ f(\mathcal{B}'^2+\mathcal{B}'') + f'\mathcal{B}' \right] \psi = \frac{M^2}{f} \psi\,.
\end{equation}
This form of the equation of motion is especially useful at zero temperature where $f = 1$. In this case, we have the Schr\"odinger-like equation
\begin{equation}
\label{eq:axionsch2}
-\psi'' + \mathcal{V}(z) \psi = M^2 \psi,
\end{equation}
where the potential $\mathcal{V}$ is defined as
\begin{equation}
\label{eq:potential}
\mathcal{V}(z) = \mathcal{B}'(z)^2 + \mathcal{B}''(z)\,.
\end{equation}
The pole of the corresponding Euclidean Green's function is obtained by imposing a Dirichlet condition for the fluctuation at the boundary while at the horizon $z_h$ the axion fluctuation must be finite. This completely specifies the eigenvalue problem to find $M^2$.

Let us now state some basic facts about the bulk axion spectrum in these theories. First, $M^2$ is real. Second, the spectrum is gapped ($M^2 > 0$) if there  is a black brane horizon in the bulk. Third, the spectrum is discrete. 

That the spectrum is purely real follows simply from the fact that Eq. \eqref{eq:axioneom} and its boundary conditions are posed as a Sturm-Liouville problem.

Now, let us analyze the mass gap. It is easy to see $M = 0$ is not in the spectrum. If $M=0$, then the equation of motion \eqref{eq:axioneom} can be easily integrated yielding two linearly independent solutions, $a \propto \, \mathrm{const}$ and $a \propto\, \int dz\,  e^{-2\mathcal{B}} f^{-1}$. The solution $a \propto \mathrm{constant}\neq 0$ is not normalizable in the UV and must be discarded. The other solution is normalizable; however, near the horizon, as $f(z) \sim -|f'(z_h)| (z_h - z)$ and $\mathcal{B} \sim \mathcal{B}(z_h)$, $a \propto \log(z-z_h) \to \infty$. Thus, the normalizable solution in the UV is not finite on the horizon. Thus, $M=0$ does not satisfy the boundary conditions and is not in spectrum if there is a horizon.

To prove that $M^2 < 0$ is not allowed we employ an argument used by Witten \cite{Witten:1998zw}. The equation of motion \eqref{eq:axioneom} can be obtained from the on shell action
\begin{equation}
\label{eq:axionaction2}
\frac{1}{32 \pi G_5\,T}  \int_0^{z_h} dz\, \mathcal{Z}(z)e^{3\mathcal{A}}\,\left[f  (\partial_z a)^2 - M^2  a^2 \right]\,.
\end{equation}
If $a$ is a normalizable solution of the equations of motion, then after integrating Eq.\ \eqref{eq:axionaction2} by parts one sees that it must vanish. Now, suppose that $M^2 <0$. Then in Eq. \eqref{eq:axionaction2} both terms are strictly positive. Thus, we must have $da/dz=0$ and $a = 0$, since the solution is normalizable. This is just the trivial solution and, thus, $M^2 < 0$ cannot be an eigenvalue of the equation of motion. Therefore, as we have already shown that $M \neq 0$, we see that $M^2>0$. This shows the existence of the mass gap.

Finally, intuitively, the spectrum must be discrete - the axion is confined into an asymptotically AdS$_5$ spacetime with a black brane deep in the bulk. The Dirichlet asymptotic boundary and the horizon work as two ``walls" that confine the axion into an infinite well, hence the discrete spectrum.

\section{Debye screening mass in strongly coupled $\mathcal{N} = 4$ SYM theory}

\label{sec:adsdebye}

\subsection{The axion spectrum}

In this section we review the holographic evaluation of the Debye mass (i.e., the smallest thermal mass associated with axion fluctuations in the bulk) in a strongly coupled $\mathcal{N} = 4$ SYM plasma by means of the gauge/gravity duality \cite{Bak:2007fk}. Since the dilaton is constant in this case the equations of motion for the dilaton and the axion fluctuations are degenerate. Also, the $\mathcal{Z}$ function is constant and one can set it to unity since one can consistently set the other bulk fields in type IIB supergravity, apart from the metric and the five-form $F_5$, to be trivial. Thus, one can simply retrieve the result from Ref.\ \cite{Csaki:1998qr} for the spectrum of a massless scalar field in a Schwarzschild $AdS_5$ background. The final result for the ground state is given by Ref. \cite{Bak:2007fk}
\begin{equation}
\label{eq:n4debye}
m_D = c\, \pi \,T
\end{equation}
where $c = 3.4041$. Since the analytical and numerical procedures used in this case will be applied with minimal changes in the next two sections, it will be useful to review here the numerical procedure used to determine the constant $c$ defined above in some detail.

For the AdS$_5$ Schwarzchild background the black brane temperature is $T = \pi R^2 z_h$. The equation of motion is given by Eq.\ (\ref{eq:axionsch}); it is useful to write it in terms of the normalized variable $u = z/z_h = \pi R^2 T z$ and the dimensionless mass $\tilde{M} = M/(\pi T)$. We need to match the solution of the equation of motion Eq.\ \eqref{eq:axionsch} with the asymptotic equation of motion near the boundary $u \to 0$
\begin{equation}
\label{eq:asyeq}
-\frac{d^2\psi}{du^2} + \mathcal{V}_{\mathrm{asy}} \psi = \tilde{M}^2 \psi,
\end{equation}
with the asymptotic potential $\mathcal{V}_{\mathrm{asy}}(u) = \mathcal{V}(u \to 0) = 15/(4u^2)$ (see Eq.\ \eqref{eq:potential}). The general solution of the asymptotic equation \eqref{eq:asyeq} is
\begin{equation}
\label{eq:asysol}
\psi(u) = C_1 \sqrt{u}[J_2(\tilde{M} u) + C_2 Y_2 (\tilde{M} u)],
\end{equation}
where $J_n$ and $Y_n$ are Bessel functions of the first and second kind, respectively, and $C_1$, $C_2$ are integration constants. Since $Y_2$ does not vanish at the boundary, we pick $J_2$ as the asymptotic solution setting $C_2 = 0$. The coefficient $C_1$ is chosen to fix the leading coefficient of the series expansion of the Bessel function to 1; then $C_1 = M^2/8$. Thus, at the boundary, the full solution $\psi$ must match the asymptotic solution
\begin{equation}
\label{eq:asysol}
\psi_{asy}(u) = \frac{8\sqrt{u}}{\tilde{M}^2} J_2(\tilde{M} u) = u^{5/2} - \frac{1}{12} \tilde{M}^2 u^{9/2} + \frac{1}{384} \tilde{M}^4 u^{13/2} + \mathcal{O}(\tilde{M}^6 u^{17/2}). 
\end{equation}

To obtain the axion spectrum numerically we use a shooting procedure. One starts with an initial value for $\tilde{M}^2$ and numerically solve the equation of motion (\ref{eq:axionsch}) imposing as boundary conditions that the solution $\psi(u)$ matches the asymptotic solution \eqref{eq:asysol} and its first derivative for some $u_0 \ll 1$. One then integrates the initial value problem up to near the horizon. When $\tilde{M}^2$ is not an exact eigenvalue $\psi(u)$ diverges at the horizon. However, $\psi (u \to 1)$ changes sign when one passes by an exact eigenvalue and, thus, one can bracket it by scanning when such sign change happens. Care must be taken to certify that one has not missed the ground state (or an excited state) by starting with values of $\tilde{M}^2$ only slightly above zero. Proceeding this way, one obtains for the ground state of the axion spectrum $\tilde{M} = m_D/(\pi T)$ the result \eqref{eq:n4debye}.

\section{Debye screening mass in Model A}
\label{sec:ihqcd}

\subsection{General IHQCD backgrounds}

The IHQCD model for $SU(N_c)$ Yang-Mills theory proposed in \cite{Gursoy:2007cb,Gursoy:2007er,Gursoy:2008bu,Gursoy:2008za,Gursoy:2009jd} corresponds to write the most general gravitational effective action involving the metric (which is dual to the energy-momentum tensor of the gauge theory) and the dilaton $\phi$ (dual to the dimension 4 scalar glueball operator ${\rm Tr} \,F^2$ in the gauge theory) with at most two derivatives in the bulk. In this model, $e^\phi$ is related to the gauge coupling. The effective 5-dimensional action in the Einstein frame for the metric and the dilaton in IHQCD is
\begin{equation}
\label{eq:kajantieaction2}
S = \frac{1}{16 \pi G_5} \int d^5x \sqrt{g} \left[\mathcal{R}- \frac{4}{3}(\partial \phi)^2 + V(\phi) \right],
\end{equation}
plus a Gibbons-Hawking boundary term, necessary to give a well posed variational problem (this term and other contributions needed in the process of holographic renormalization \cite{Bianchi:2001kw,Skenderis:2002wp} do not affect our discussion and are, thus, dropped altogether). The potential $V(\phi)$ is assumed to contain part of the sub-critical 5d string theory contributions to the effective action. The metric background (in the Einstein frame) is written in the conformal gauge in the usual form
\begin{equation}
\label{eq:kajantiebg}
ds^2 = b(z)^2 \left[f(z)d\tau^2 + d\vec{x}^2+\frac{dz^2}{f(z)} \right],
\end{equation}
while the dilaton is assumed to depend only upon the radial coordinate $z$, $\phi = \phi(z)$ and $\tau$ is a periodic coordinate with period $1/T$. Comparing with Eq.\ \eqref{eq:bgmetric}, one sees that $b(z) = e^{\mathcal{A}(z)}$. The Einstein's and scalar equations of motion that follow from extremizing \eqref{eq:kajantieaction2} are
\begin{align}
\label{eq:kajantieeom1}
\frac{f''}{f'} + 3 \frac{b'}{b} & = 0, \nonumber \\
6 \frac{b'^2}{b^2} - 3\frac{b''}{b} & = \frac{4}{3} \phi'^2 \quad \quad \mathrm{and} \\
6 \frac{b'^2}{b^2} + 3\frac{b''}{b} + 3 \frac{b'}{b} \frac{f'}{f} & = \frac{b^2}{f} V, \nonumber
\end{align}
where the prime indicates differentiation with respect to $z$. The equation of motion for $\phi$ is a combination of the previous equations - as usual for Einstein's equations there is some redundancy in the equations of motion (due to Bianchi's identity).

\subsection{An exact solution - model A}\label{kajantiemodel}

An analytical solution \cite{Gursoy:2007er} of the equations of motion \eqref{eq:kajantieeom1} is given by trying the following Ansatz 
\begin{equation}
\label{eq:kajantieb}
b(z) = \frac{R}{z} e^{-\frac{1}{3} \Lambda^2 z^2},
\end{equation}
where $\Lambda$ is an infrared scale of the order of $\Lambda_{QCD}$. Defining the dimensionless variable $y \equiv \Lambda z$ and $\lambda = e^{\phi}$, one can integrate the equations of motion to find \cite{Kajantie:2011nx}
\begin{equation}
\label{eq:kajantielambda}
\frac{\lambda(y)}{\lambda_0} = \exp \left(\frac{\sqrt{\frac{9}{y^2}+4} \, y \left(2 \sqrt{4 y^2+9} \, y+9 \sinh ^{-1}\left(\frac{2 y}{3}\right)\right)}{8 \sqrt{4 y^2+9}}\right)
\end{equation}
where $\lambda_0 = \lambda(0)$. Also, the horizon function is given by
\begin{equation}
\label{eq:kajantiehorizon}
f(y,y_h) = 1 -\frac{(y^2-1)e^{y^2} +1}{1+e^{y_h^2}(y_h^2-1)}\,.
\end{equation}
Finally, the dilaton potential for this solution is given by
\begin{equation}
\label{eq:kajantiepot}
V(y,y_h) = \frac{12}{R^2} e^{\frac{2y^2}{3}} \left[ \left(\frac{1}{3} y^4 + \frac{5}{6} y^2 + 1) \right)f(y,y_h) - \left(\frac{1}{2} + \frac{1}{3} y^2 \right) \frac{y}{2} \frac{\partial f}{\partial y} (y,y_h) \right].
\end{equation}
Note that the potential depends explicitly on the temperature via the position of the horizon $y_h$. This is not going to be the case in the other type of models considered in \ref{ref:Bclass}.

\subsubsection{Thermodynamics}

The temperature of the thermal bath is given by the Hawking temperature of the black brane
\begin{equation}
\label{eq:kajantietemp}
T = \frac{|f'(z_h)|}{4\pi} = \frac{\Lambda}{2\pi} \frac{y_h^3}{y_h^2-1+ e^{-y_h^2}}\,.
\end{equation}
The entropy density is given by the Bekenstein-Hawking formula, which yields
\begin{equation}
\label{eq:kajantieent}
s = \frac{b(z_h)^3}{4 G_5} =  \frac{R^3 \Lambda^3}{4 G_5} \frac{e^{-y_h^2}}{y_h^{3}}\,.
\end{equation}
Moreover, the pressure follows from $s=\partial p/\partial T$
\begin{equation}
\label{eq:kajantiepressure}
p(y_h) = - \int_{y_h}^{\infty} s(T(x)) \frac{dT(x)}{dx}
\end{equation}
and the energy density is given by $\epsilon = sT-p$.

\begin{figure}
\centering
  \includegraphics[width=.5\linewidth]{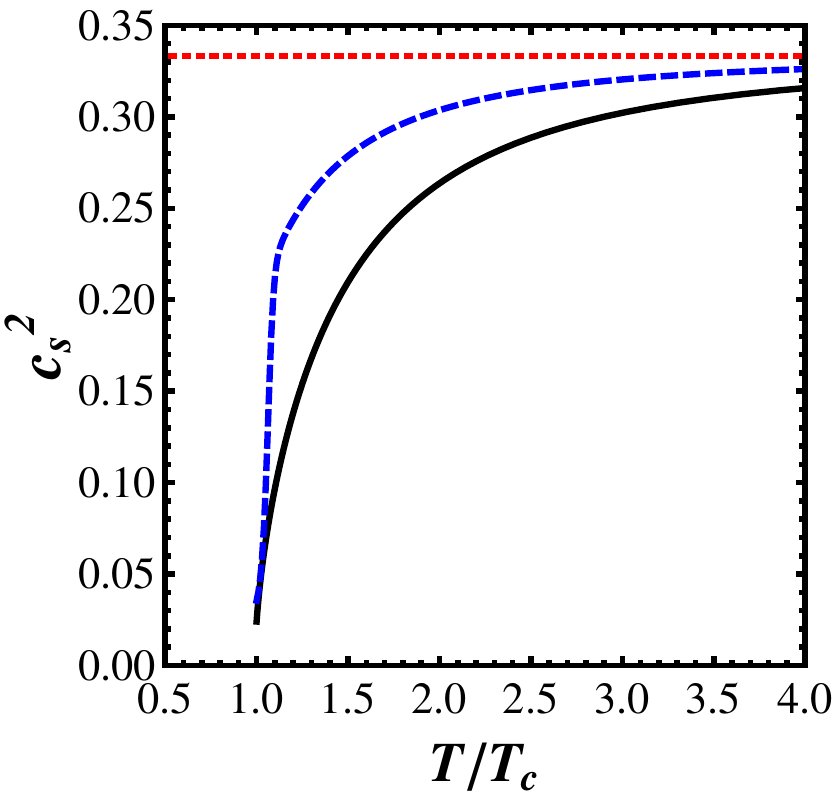}
  \caption{(Color online) Speed of sound squared $c_s^2$ of the plasma as a function of $T/T_c$, where $T_c$ denotes the critical temperature for a deconfining first order transition. The solid black line is the result for the particular IHQCD model studied (see \ref{kajantiemodel}), the dashed blue line corresponds to lattice results from \cite{Panero:2009tv} for an SU(3) Yang-Mills plasma while the horizontal red line gives the result for a conformal plasma, $c_s^2 = 1/3$.}
  \label{fig:cs2kajantie}      
\end{figure}

The temperature function $T(y_h)$ has a minimum for $T_{\mathrm{min}} = 0.396 \Lambda$, $y^{\mathrm{min}}_h = 1.466$. For $T< T_{min}$, there is no possible black brane solution and the system is in a thermal gas phase. However, for $T_{\mathrm{min}} < T < T_c$, where $T_c = 0.400 \Lambda$ is reached at $y^{c}_h = 1.299$, albeit there is a black brane solution, the pressure is negative - this signs that the thermal plasma is in a metastable phase. For $T>T_c$ (thus, $y_h<y_h^c$), we have a deconfined thermal plasma state. Since the entropy density has a discontinuity at $T=T_c$, the transition is of first order. It is possible to explicitly write the equation of state of the system in terms of the speed of sound squared
\begin{equation}
\label{eq:kajantiecs2}
c_s^2=\frac{d \log T}{d \log s} = \frac{1}{3+y_h^2} \frac{3-y_h^2-(3+2y_h^2)e^{-y_h^2}}{y_h^2-1+e^{-y_h^2}}.
\end{equation}
In Fig.\ \ref{fig:cs2kajantie} we compare the equation of state of this model, given by \eqref{eq:kajantietemp} and \eqref{eq:kajantiecs2}, with lattice results for a pure glue SU(3) Yang-Mills plasma \cite{Panero:2009tv}. We see that this gravity dual provides a reasonable qualitative description of the equation of state of a pure glue plasma, more so considering its relative simplicity and the fact that it is an analytical solution of the Einstein+scalar equations of motion. However, it must be noted that this simple realization of IHQCD does not describe lattice data quantitatively between $T/T_c=1.1-2.5$\footnote{This can be remedied by choosing an appropriate dilaton potential and, as shown in \cite{Gursoy:2007cb,Gursoy:2007er,Gursoy:2008bu,Gursoy:2008za,Gursoy:2009jd}, a good quantitative agreement with pure glue lattice QCD thermodynamics in this temperature range can be achieved.}. 

\subsubsection{Polyakov loop}
\label{sec:polyloop}

An interesting quantity to compute in this non-conformal model is the expectation value of the Polyakov loop operator \cite{polyakov1,polyakov2,polyakov3,McLerran:1980pk}
\begin{equation}
\label{eq:polyop}
\hat{L}(\vec{x}) = \frac{1}{N_c} \mathrm{Tr} \, P \exp \left(i \int_0^{\beta} \hat{A}_0 (\tau, \vec{x}) d\tau \right),
\end{equation}
where $P$ indicates path-ordering and the trace is in the fundamental representation. Holographically, the evaluation of the Polyakov loop in a thermal gauge theory in the imaginary time formalism corresponds to calculating the classical worldsheet action for a {\it straight} string in the bulk stretching from the conformal boundary to the horizon. This string worldsheet wraps the imaginary time circle $S^1$ (for details of the holographic computation of the Polyakov and Wilson loops in this context, see \cite{Maldacena:1998im,Brandhuber:1998bs,Rey:1998bq,Noronha:2009ud,Noronha:2010hb,Noronha:2009da,Finazzo:2013rqy}). At strong coupling and large $N_c$, the norm of the expectation value of the Polyakov loop operator (\ref{eq:polyop}) is given by
\begin{equation}
\label{eq:holopoly}
|\langle \hat{L}\rangle| \sim e^{- F_Q/T} \sim e^{-S_{NG}},
\end{equation}
where $F_Q$ is the difference in the free energy of the thermal bath due to the inclusion of a single probe heavy quark in the system, and $S_{NG}$ is the (Euclidean) Nambu-Goto action for the string worldsheet
\begin{equation}
\label{eq:nambugoto}
S_{NG} = \frac{1}{2\pi \alpha'} \int d^2 \sigma \sqrt{\det\left(g^s_{\mu \nu} X^{\mu}_{a} X^{\nu}_{b} \right)},
\end{equation}
where $\alpha'=\ell_s^2$, $\ell_s$ is the string length, $X^{\mu}_{a}$ are the embedding functions of the string worldsheet in the target space-time, and $g^s_{\mu \nu}$ is the metric in the string frame - since this background comes from a 5 dimensional non-critical string theory, $g^s_{\mu \nu} = \lambda^{4/3} g_{\mu \nu}$, where $g_{\mu \nu}$ is the metric in the Einstein frame \cite{Gursoy:2007cb,Gursoy:2007er}. The indices $\mu,\nu = 0,1,2,3,4$ are spacetime indices and $a,b = \sigma, \tau$ are indices for the string worldsheet coordinates. Evaluating the worldsheet specified above with the background \eqref{eq:kajantiebg} one can see that
\begin{equation}
\label{eq:FQnreg}
F_Q = \frac{1}{2\pi \alpha'} \int^{y_h}_0 dy \, \sqrt{g_{00} g_{zz}} = \frac{1}{2\pi \alpha'} \int^{y_h}_0 dy \, b^2_s(y), 
\end{equation}
where $b_s \equiv \lambda^{2/3} b$. As expected, the bare heavy quark free energy is UV divergent and must be regularized. To regularize it, we use a temperature independent subtraction 
\begin{equation}
\label{eq:FQregulator}
\bar{F}_Q = \frac{1}{2\pi \alpha'} \int_0^{y_h^c}dy\, (b_s^{(0)}(y))^2,
\end{equation}
where $b_s^{(0)}(y)$ is the vacuum form of $b_s(y)$. The regularized free energy is then $F^{reg}_Q = F_Q - \bar{F}_Q$. For the geometry in question
\begin{equation}
\label{eq:FQreg}
\frac{F^{reg}_Q T_c}{\sigma} =  -\frac{T_c}{\Lambda b_s^2(y_{min})} \int^{y^c_h}_{y_h} dy\, b_s^2 (y)
\end{equation}
where, to facilitate the comparison with lattice results, we normalized the heavy quark free energy by the holographically computed string tension $\sigma$ (associated with the area law for the rectangular Wilson loop in the vacuum) and by the critical temperature $T_c$. The holographic string tension in IHQCD is generally given by $\sigma = R^2 \Lambda^2 b_s^2(y_{min})/(2\pi \alpha')$ \cite{Gursoy:2007er}, where $y_{min}$ denotes the location of the minimum of the U-shaped Nambu-Goto string profile in the bulk used in the calculation of the rectangular Wilson loop for asymptotically large separations. 

Note that the Polyakov loop computed on the lattice depends on the choice of renormalization scheme since the heavy quark bare free energy is divergent in the continuum limit (one needs to subtract the divergent part and fix the renormalization constant). In the calculations of Ref.\ \cite{Mykkanen:2012ri} this constant term was set to zero. Clearly, any other value for the constant would be fine and the scheme dependence just corresponds to adding an additive constant in the free energy of the renormalized Polyakov loop \footnote{We thank M.~Panero for discussions about this point.}. In this paper we chose to compare the free energy difference $\Delta F_Q T_c/ \sigma = (F_Q(T)-F_Q(2T_c))T_c/\sigma$ as a function of $T/T_c$ computed in the model with the one found on the lattice (note that this still corresponds to choosing a scheme in which the free energy difference vanishes at $2T_c$).

\begin{figure}
\centering
  \includegraphics[width=.5\linewidth]{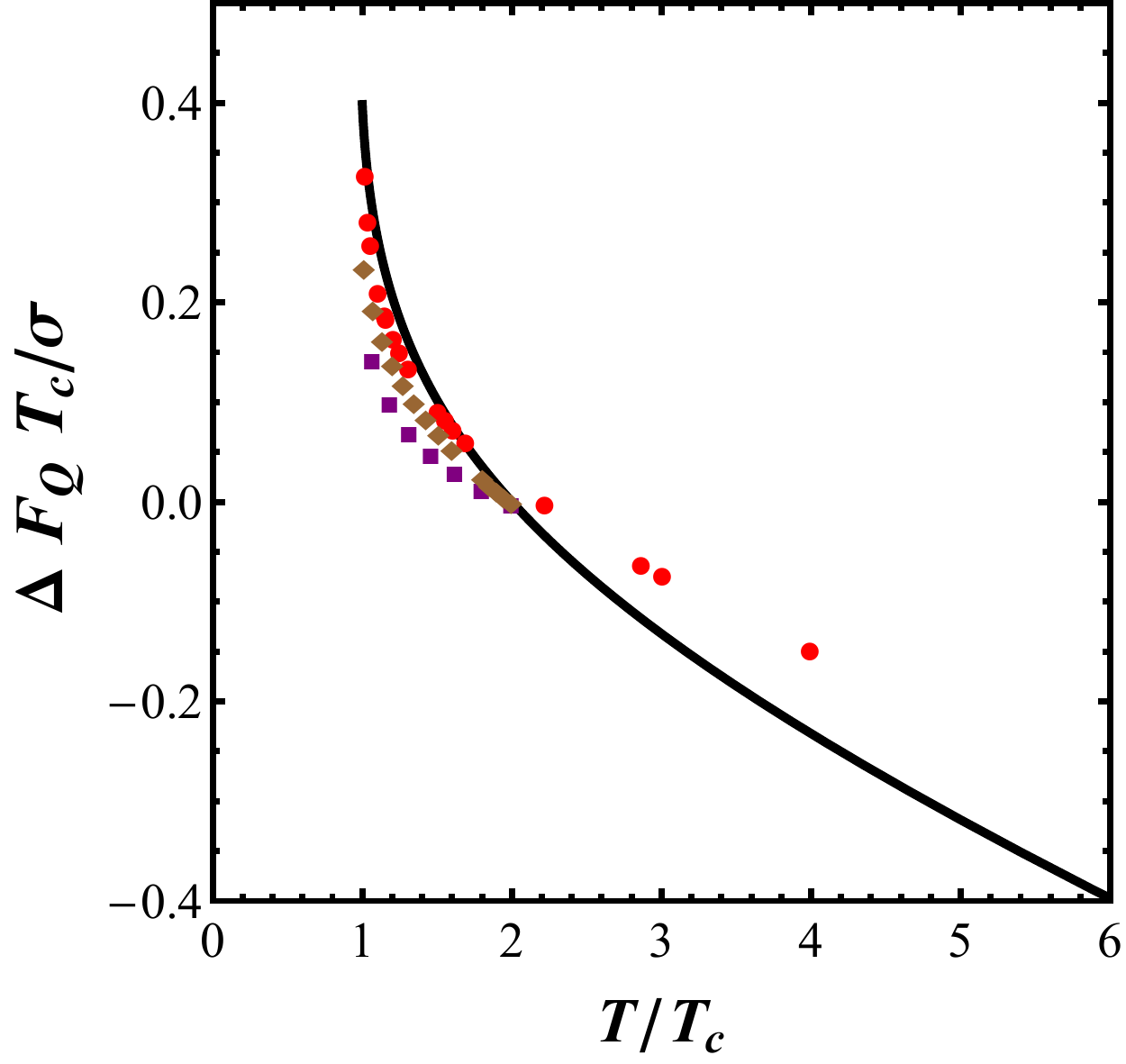}
  \caption{(Color online) $\Delta F_Q T_c/ \sigma = (F_Q(T)-F_Q(2T_c))T_c/\sigma$ as a function of $T/T_c$ for model A (solid black line) defined in \ref{kajantiemodel} and for $SU(N_c)$ Yang-Mills \cite{Mykkanen:2012ri} with $N_c =$ 3 (red circles), 4 (purple squares), and 5 (brown diamonds).}
  \label{fig:FQgraf}      
\end{figure}

We compare in Fig.\ \ref{fig:FQgraf} the holographic result \eqref{eq:FQreg} with the lattice results for the $SU(N_c)$ Yang-Mills lattice data with different number of colors from \cite{Mykkanen:2012ri}. One can see that even though the thermodynamics of the simple IHQCD model only reproduces qualitatively the lattice data, the holographic result for $F_Q$ gives a reasonable description of the lattice data for $T_c <T <2 T_c$. Moreover, even though holographic models ought to be valid only for large $N_c$, reasonable agreement is seen even for $N_c = 3$.

\subsubsection{Debye screening mass}\label{debyesectionmodelA}

Let us begin by studying the bulk axion spectrum at $T=0$ (i.e., we set $f=1$). The first step is to discuss the function $\mathcal{Z}$ in the action for the axion fluctuations \eqref{eq:axionaction}, which represents a partial resummation of higher order forms coming from 5-dimensional sub-critical string theory \cite{Gursoy:2007cb,Gursoy:2007er}. In the UV, $Z(\lambda) \sim \mathrm{const}$ while in the IR $(\lambda) \propto \lambda^4$ to ensure glueball universality. We will use the following standard IHQCD parametrization that interpolates between these two cases \cite{Gursoy:2012bt}
\begin{equation}
\label{eq:Zparam}
Z(\lambda) = c_0 + c_4 \lambda^4,
\end{equation}
where $c_0$ and $c_4$ are constants. By a suitable normalization of the action one can set $c_0=1$. To study the dependence of the results with $c_4$, we choose three values for it spanning a large range of values for this coefficient: 0.1, 1, and 10.

The numerical procedure to find the spectrum is the same as the one described in \ref{sec:adsdebye}. For the vacuum case we consider the Schr\"odinger equation \eqref{eq:axionsch2} and the asymptotic potential in the UV, including the first subleading correction in $1/y$, which gives
\begin{equation}
\label{eq:kanjantieasy}
\mathcal{V}(y) = \frac{15}{4y^2} - \frac{9 \sqrt{2} c_4}{(1+c_4)y} + \mathcal{O}(1)\,.
\end{equation}
The asymptotic equation (including the subleading term) can be solved analytically and the linearly independent solutions are Whittaker functions $M_{\kappa,\mu}$ and $W_{\kappa, \mu}$ \cite{gradshteyn}. If we consider only the leading term in $1/y$, these solutions reduce to the Bessel functions found in \ref{sec:adsdebye}.  The normalized near boundary series expansion, including the subleading term in \eqref{eq:kanjantieasy}, is given by
\begin{equation}
\label{eq:kajantiebound}
\psi (y) = y^{1/2}\left(y - \frac{9\sqrt{2}c_4}{5} \frac{y^{2}}{1+c_4} + \cdots\right)\,.
\end{equation}

Using the shooting method to solve the eigenvalue problem, we obtain the results shown in Table \ref{tab:axion}. One can see that glueball mass associated with the bulk axion in the vacuum is quite insensitive to the choice of $c_4$ and $m_{J^{PC}=0^{-+}} \sim 3 \Lambda$. This value is also comparable with the corresponding results for the lightest $J^{PC}=0^{++}$ and $J^{PC} = 2^{++}$ glueballs in this model, $m_{J^{PC}=0^{++}} \approx 2.5 \Lambda$ and $m_{J^{PC}=2^{++}} = \sqrt{8} \Lambda \sim 2.2 \Lambda$ \cite{Kajantie:2011nx}.
\begin{table}
\begin{centering}
\begin{tabular}{cc}
\hline
$c_4$    & $m_{\mathrm{axion}}/\Lambda$ \\
\hline
0.1 & 3.0433 \\
1.0 & 2.996 \\
10.0 & 2.986 \\
\hline
\end{tabular}
 \caption{\label{tab:axion} Glueball mass $m_{J^{PC}=0^{-+}}$ associated with the bulk axion at $T=0$ for some choices of $c_4$ computed using the model in \ref{kajantiemodel}.}
\end{centering}
\end{table}

\begin{figure}
\centering
  \includegraphics[width=.5\linewidth]{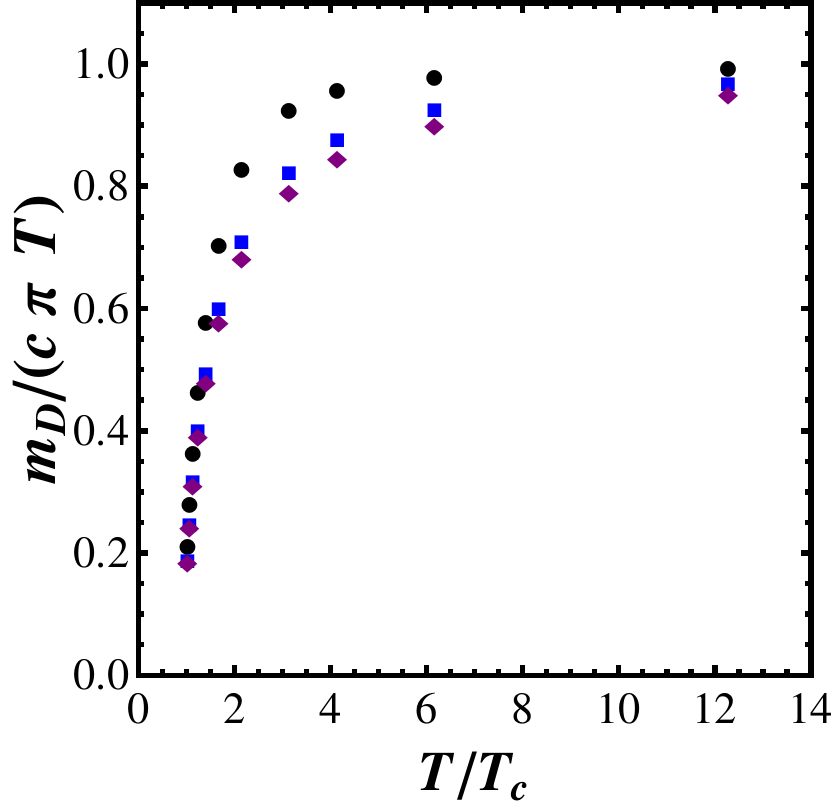}
  \caption{(Color online) Debye screening mass $m_D$ for the simplified IHQCD model discussed in \ref{kajantiemodel}, normalized by the $\mathcal{N}=4$ SYM Debye mass at strong coupling $c \pi T$ with $c=3.4041$. We present the results for $c_4 = $ 0.1 (black circles), 1 (blue squares) and 10 (purple diamonds).}
  \label{fig:debyemasskajantie}      
\end{figure}

Let us now proceed to extract the Debye screening mass in this model. Consider now the background at nonzero temperature. The equation of motion to solve is now of the form \eqref{eq:axionsch}. The asymptotic solution is the same as in the $T=0$ case since $f\to1$ for $y\to 0$. We use the same choices for $c_4$ employed in the preceding calculation. Our results can be found in Fig.\ \ref{fig:debyemasskajantie}. Since at high temperatures $T \gg \Lambda$ the geometry of the gravity dual simplifies to $AdS_5$, one must have $m_D(T \gg \lambda) \to c\, \pi T$ with $c=3.4041$ as shown in Section \ref{sec:adsdebye}. Thus, our results for $m_D$ are normalized by $c\,\pi T$.

One can see that the results are somewhat insensitive to the choice of $c_4$ as long as $c_4 \gtrsim 1$. Also, we note that for $T \to T_c^+$, $m_D/(c \pi T) \sim 0.18$, which is nearly independent of $c_4$ - the Debye mass has a discontinuity at $T=T_c$. As expected, for increasing temperature, the plasma becomes more and more screened - $m_D$ is monotonically increasing with $T$ until it reaches its conformal value. 

Ref.\ \cite{Hoyos:2011uh} computed the thermal screening lengths for an $\mathcal{N}=2^*$ plasma, which is non-conformal deformation of the $\mathcal{N}=4$ SYM plasma obtained by giving a mass $\mu$ to the adjoint scalars and fermions \cite{Pilch:2000ue,Buchel:2003ah,Buchel:2007vy}. Using this top-down non-conformal construction \cite{Hoyos:2011uh} also obtained that $m_D/T$ (computed from the axion fluctuations) becomes smaller than its conformal value at low temperatures when $\mu/T>1$. However, this theory does not possess a finite temperature phase transition and, thus, the discontinuity in $m_D/T$ at $T_c$ found here is a new feature brought in by the non-conformal plasmas constructed within IHQCD.    

\section{B class of models - Overview}
\label{ref:Bclass}

In this section we shall describe a second class (Model B) of strongly coupled non-Abelian plasmas with gravity duals described by Einsten+scalar actions \cite{Gubser:2008ny,Gubser:2008yx,Gubser:2008sz} (see also \cite{Noronha:2009ud,Noronha:2010hb,Finazzo:2013efa}) built in order to reproduce some of the thermodynamic results obtained on the lattice at zero baryon chemical potential. Even though the bulk fields are the same as in the previous section, in these models the scalar field corresponds to a relevant operator in the UV. 

The interpretation put forward in \cite{Gubser:2008yx} is that since these gravity models cannot truly describe perturbative QCD physics in the UV, one must choose an intermediate semi-hard scale at which asymptotic freedom is replaced by conformal invariance. In fact, given that the scaling dimension $\Delta$ of the glueball operator ${\rm Tr}\,F^2$ is not a protected quantity in QCD and it becomes smaller than 4 towards the IR, this semi-hard scale may be used to define the range of applicability of this effective holographic model in this context. This implies that, in general, these models should not be used at high temperatures where asymptotic freedom becomes dominant. However, as shown in \cite{Finazzo:2013efa}, these non-conformal bottom-up models are able to describe not only the equilibrium quantities found on the lattice but also the temperature dependence of some nontrivial transport coefficients such as the electrical conductivity recently computed on the lattice \cite{Amato:2013naa}. Moreover, these models also give valuable insight into the energy loss experienced by heavy (and also light quarks) in the QGP near the crossover phase transition \cite{Ficnar:2010rn,Ficnar:2011yj,Ficnar:2012yu}. Therefore, we believe that it is relevant to consider these constructions here as well and investigate the temperature dependence of $m_D/T$ in these models. We shall see that by carefully choosing the scalar potential one can obtain a much better quantitative description of the thermodynamics of pure glue as well as that of QCD with light dynamical flavors found on the lattice\footnote{{We note that the models considered here do not have the correct bulk degrees of freedom to fully describe the physics associated with chiral symmetry breaking. See Refs. \cite{Li:2013oda,Li:2014dsa} for a model which describes chiral symmetry breaking in this class of Einstein + scalar models by including a second scalar field, following the spirit of the KKSS model \cite{Karch:2006pv}.}}.   

\subsection{Bulk action}

Even though the bulk action that defines these models is the same as that studied in \ref{sec:ihqcd}, we find it convenient to follow the convention of Ref.\ \cite{Gubser:2008ny} (compare the dilaton normalization in Eq.\ \eqref{eq:kajantieaction2} with the one below) where the Einstein+scalar action is
\begin{equation}
\label{eq:gubseraction}
S=\frac{1}{2 \kappa_5^2} \int d^5x\, \sqrt{g} \left[\mathcal{R} - \frac{1}{2} (\partial_{\mu}  \phi)(\partial^{\mu}  \phi) - V(\phi) \right],
\end{equation}
where $k_5^2 = 8 \pi G_5$. The scalar field in this action is related to the dilaton in Model A \eqref{eq:kajantieaction2} by a factor of $\sqrt{3/8}$.  The potential $V(\phi)$ is chosen in such a way that the thermodynamic properties of the model \eqref{eq:gubseraction} mimic the ones desired from the gauge theory - in the next subsections we will describe simple choices of $V(\phi)$ which achieve this task. The desired solutions of Eq.\ \eqref{eq:gubseraction} must be asymptotically $AdS_5$ for the boundary gauge theory to have a UV fixed point. The potential $V(\phi)$ is chosen in order to interpolate between a free massive scalar field (plus cosmological constant term) near the boundary, $V(\phi) \sim -12/R^2 + m^2 \phi^2/2$, and a potential which yields the Chamblin-Reall solution \cite{Chamblin:1999ya} deep in the bulk, $V(\phi) = V_0 e^{\gamma \phi}$, with $\gamma < 0$.

\subsection{Metric Ansatz}

As we wish to study the gauge theory at finite temperature, the solution also must contain a black brane in the bulk. We also want translation symmetry in the gauge theory and rotational SO(3) symmetry in the spatial directions but not the full Lorentz SO(3,1) symmetry since the at nonzero temperature the thermal gauge theory is not invariant by Lorentz boosts. An Ansatz which is able to satisfy these requirements, called here the Gubser gauge \cite{Gubser:2008ny}, is
\begin{equation}
\label{eq:gubseransatz}
ds^2 = e^{2A} (h \ d\tau^2 + d\vec{x}^{\,2}) + e^{2B} \frac{d\phi^2}{h},
\end{equation}
where the holographic radial coordinate is given by the scalar field $\phi$ itself. We require that $A$, $B$, and $h$ are only functions of $\phi$, i.e., $A(\phi)$, $B(\phi)$, and $h(\phi)$. The asymptotically $AdS_5$ boundary is recovered when $\phi \to 0$. This choice, as shown in \cite{Gubser:2008ny}, is convenient to solve the equations of motion for the action \eqref{eq:gubseraction}. However, this gauge choice is not very useful for analyzing the glueball spectra or studying Wilson and Polyakov loops. For these purposes, it is convenient to go back to conformal gauge. We discuss this point in more detail in Appendix \ref{sec:appendix-gauge}.

\subsection{The equations of motion - general case}

It is possible to write a ``master" equation that yields all the metric functions in the Ansatz \eqref{eq:gubseransatz} in terms of a single ordinary first order differential equation \cite{Gubser:2008ny}. The equations of motion derived from the action \eqref{eq:gubseraction} are the Einstein's equations
\begin{equation}
\label{eq:einsteineom}
\mathcal{R}_{\mu\nu}-\frac{g_{\mu\nu}}{2}\mathcal{R} = 8 \pi G_5 T_{\mu \nu},
\end{equation}
where $T_{\mu \nu}$ is the stress-energy tensor for the scalar field. The equation of motion for the scalar field $\phi$ is
\begin{equation}
\label{eq:scalareom}
\nabla_{\mu} \nabla^{\mu} \phi - V' = 0,
\end{equation}
where $\nabla$ indicates the covariant derivative and $V'=dV/d\phi$ (in this section, primes will always indicate derivatives with respect to $\phi$). With the Ansatz \eqref{eq:gubseransatz}, one can see that the equation of motion for the $\tau\tau$ component is
\begin{equation}
\label{eq:gtteom}
2 e^{2B} V + 6 A' h' + h (24 A'^2 -12B'A'+12A''+1) = 0 
\end{equation}
while for the $xx$ the equation of motion is
\begin{equation}
\label{eq:gxxeom}
2 e^{2B} V + 14 A' h' - 2B' h' + 2h'' + h (24 A'^2 -12B'A'+12A''+1) = 0\,.
\end{equation}
The common term in parenthesis can be eliminated from both equations, which yields
\begin{equation}
\label{eq:gubsereom1}
h''+(4 A'-B')h'=0\,.
\end{equation}
The $G_{\phi \phi}$ equation of motion is
\begin{equation}
\label{eq:gubsereom2}
6A'h'+h(24A'^2 -1) + 2 Ve^{2B} = 0\,.
\end{equation}
Using the $G_{\tau\tau}$ equation of motion \eqref{eq:gtteom} to eliminate $24 A'^2$ from Eq.\ \eqref{eq:gtteom} we obtain
\begin{equation}
\label{eq:gubsereom3}
A'' - A' B' + \frac{1}{6} = 0\,.
\end{equation}
The last equation of motion is given by the scalar equation \eqref{eq:scalareom},
\begin{equation}
\label{eq:gubsereom4}
(4 A' - B') + \frac{h'}{h} - \frac{e^{2B}}{h} V' = 0.
\end{equation}
We use the set consisting of Eqs.\ \eqref{eq:gubsereom1} to \eqref{eq:gubsereom4} as our equations of motion. These equations are not completely independent due to Bianchi's identity. In this case, the derivative of Eq.\ \eqref{eq:gubsereom3} follows from the derivative of the other equations of motion and one can use any subset of three equations among these to obtain the full geometry.

\subsection{Zero temperature master equation}
\label{sec:T0master}

We start by describing zero temperature solutions. With a vacuum solution at hand, one can proceed to explore the properties of the $T=0$ strongly coupled non-Abelian gauge theory with gravity dual given by Eq.\ \eqref{eq:gubseraction}. Although this class of models was built primarily in order to reproduce the thermodynamics of QCD near the crossover phase transition \cite{Aoki:2006we}, in Appendix \ref{sec:appendix-spectra} we show that the glueball spectra is reasonably described by a confining, zero temperature version of these models.

When $T=0$, the boundary gauge theory has full Lorentz invariance and, thus, we set $h=1$ in \eqref{eq:gubseransatz}
\begin{equation}
\label{eq:gubseransatzt0}
ds^2 = e^{2A} (d\tau^2 + d\vec{x}^2) + e^{2B} d\phi^2
\end{equation}
where $\tau$ is the Euclidean time. The equation of motion \eqref{eq:gubsereom1} is identically satisfied when $h=1$. The remaining equations of motion \eqref{eq:gubsereom2}, \eqref{eq:gubsereom3}, and \eqref{eq:gubsereom4} simplify to
\begin{equation}
\label{eq:eomT01}
A'' - A'B' + \frac{1}{6} = 0,
\end{equation}
\begin{equation}
\label{eq:eomT02}
24(A')^2 - 1 + 2 e^{2B} V = 0 \quad \quad \mathrm{and}
\end{equation}
\begin{equation}
\label{eq:eomT03}
4A' - B' - e^{2B} V' = 0\,.
\end{equation}

Now, following the procedure used in Ref. \cite{Gubser:2008ny} for the $T \neq 0$ case, our goal here is to obtain a first order master equation for $G(\phi) \equiv A'(\phi)$. Then, one can integrate $G$ to obtain $A$ and the remaining metric function $B$. Combining Eqs.\ \eqref{eq:eomT02} and \eqref{eq:eomT03}, we arrive at
\begin{equation}
\label{eq:masterT01}
\frac{V}{V'} = \frac{-8G + 2 B'}{24 G^2 -1}\,.
\end{equation}
We can now use Eq.\ \eqref{eq:eomT01} to eliminate $B'$ from this equation and find the master equation at $T=0$
\begin{equation}
\label{eq:masterT0}
G + \frac{V}{3V'} = - \frac{6 G' G}{24G^2-6G'-1}\,.
\end{equation}
This is a first order ordinary differential equation for $G = A'$ for a given potential $V(\phi)$. To solve it, we have to specify a boundary condition for $G(\phi)$. Since all the potentials we shall consider have the IR $(\phi \to \infty$) asymptotic $V(\phi) \propto e^{\gamma \phi}$, we see that for $\phi \to \infty$, $V/3V' = 1/(3 \gamma)$. Thus, Eq.\ \eqref{eq:masterT0} implies that when $\phi \to \infty$ one must have
\begin{equation}
\label{eq:boundaryT0}
G (\phi \to \infty) = -\frac{1}{3 \gamma}\,.
\end{equation}

\subsection{Finite temperature master equation}

The procedure for extracting a master equation for the finite temperature case was explained in detail in \cite{Gubser:2008ny} and we shall not repeat it here. One can show that this master equation is 
\begin{equation}
\label{eq:masterT}
\frac{G'}{G+\frac{V}{3V'}} = \frac{d}{d\phi} \log \left[ \frac{G'}{G} + \frac{1}{6G} - 4G - \frac{G'}{G+\frac{V}{3V'}} \right].
\end{equation}

Let us now discuss the boundary conditions for the master equation \eqref{eq:masterT}. First, we  require that $h(\phi)$ has a simple zero at $\phi = \phi_h$, which is the radial position of the event horizon. Thus, $h(\phi_h) = 0$ but $h'(\phi_0) \neq 0$ so that for $\phi \lesssim \phi_h$, $h(\phi) \approx -h'(\phi_h) (\phi - \phi_h)$. Therefore, from Eqs.\ \eqref{eq:gubsereom2} and \eqref{eq:gubsereom3} one obtains the constraints
\begin{equation}
\label{eq:constraint1}
V(\phi_h) = -3 e^{-2B(\phi_h)} G(\phi_h) h'(\phi_h) \quad \quad \mathrm{and}
\end{equation}
\begin{equation}
\label{eq:constraint2}
V'(\phi_h) = e^{-2B(\phi_h)} h'(\phi_h).
\end{equation}
Thus, near the horizon one may expand $G + V/(3V')$ in a series around $\phi=\phi_h$
\begin{equation}
\label{eq:Gseries}
G(\phi) = -\frac{1}{3} \frac{V(\phi_h}{V'(\phi_h)} + \frac{1}{6} \left( \frac{V(\phi_h) V'(\phi_h)}{V'(\phi_h)^2} -1 \right)(\phi-\phi_h) + O[(\phi-\phi_h)^2].
\end{equation}
By fixing the position of the horizon $\phi_h$ we may use the series solution \eqref{eq:Gseries} to obtain $G(\phi)$ near the horizon, at $\tilde{\phi} = \phi_h - \delta \phi$, for $\delta \phi \ll \phi_h$, and then integrate numerically from $\phi = \tilde{\phi}$ out to $\phi = 0$ using the series values for $G(\tilde{\phi})$ and $G'(\tilde{\phi})$ as boundary conditions.

\subsection{Geometry asymptotics}

As mentioned above, the potential near the boundary ($\phi \to 0$) is given by
\begin{equation}
\label{eq:asypot}
V(\phi) \sim -\frac{12}{R^2} + \frac{m^2}{2} \phi^2\,.
\end{equation}
The UV scaling dimension $\Delta$ of the gauge theory operator associated with $\phi$ is determined by the larger root of
\begin{equation}
\label{eq:delta}
\Delta(\Delta-4) = m^2 R^2\,.
\end{equation}
In the coordinate system \eqref{eq:gubseransatz}, the asymptotic $AdS_5$ geometry ($\phi \to 0$) is given by
\begin{equation}
\label{eq:Aasy}
A(\phi) = \frac{\log \phi}{\Delta -4} + O(1) \quad \quad \mathrm{and}
\end{equation}
\begin{equation}
\label{eq:Basy}
B(\phi) = -\log \phi + O(1),
\end{equation}
with $h(\phi\to 0) \to 1$. This also fixes the asymptotic behavior $G(\phi \to 0) \sim 1/(\phi(\Delta-4))$.

\subsection{Obtaining the geometry and the thermodynamics}
\label{sec:fullgeo}

With the boundary conditions fixed and with the asymptotic behavior defined above one can obtain the full metric from $G(\phi)$. First, one can see that
\begin{equation}
\label{eq:Afnc}
A(\phi) = A_h + \int_{\phi_h}^{\phi} d\tilde{\phi} \, G(\tilde{\phi}),
\end{equation}
where $A_h=A(\phi_h)$ is the integration constant. Since near the boundary $A$ behaves as in Eq.\ \eqref{eq:Aasy}, one can obtain the integration constant $A_h$
\begin{equation}
\label{eq:Aconst}
A_h = \frac{\log \phi_h}{\Delta -4} + \int_0^{\phi_h} d\phi \, \left[ G(\phi) - \frac{1}{(\Delta-4)\phi} \right].
\end{equation}

Now, let us also evaluate $B(\phi)$ and $h(\phi)$. One can solve Eq.\ \eqref{eq:gubsereom3}  for $B'$ in terms of $G$ to obtain
\begin{equation}
\label{eq:Bfnc}
B(\phi) = B_h + \int_0^{\phi_h} d\phi \, \left[ \frac{G'(\phi)}{G(\phi)} + \frac{1}{6 G(\phi)} \right].
\end{equation}
with $B_h = B(\phi_h)$ being an integration constant, which we will determine in the end of this subsection. Also, given that $A$ and $B$ are known, one can integrate Eq.\ \eqref{eq:gubsereom1} to obtain
\begin{equation}
\label{eq:hfnc}
h(\phi) = h_0 + h_1 \int^{\phi}_{\phi_h} d\tilde{\phi} \, e^{-4A(\tilde\phi) + B(\tilde{\phi})},
\end{equation}
where $h_0$ and $h_1$ are integration constants. To determine them, remember that $h(\phi \to 0) = 1$ and $h(\phi_h) = 0$ so that
\begin{equation}
\label{eq:hconst}
h_0 = 0 \quad \quad \mathrm{and} \quad \quad h_1 = \frac{1}{\int_{\phi_h}^0 d\phi \, e^{-4A(\phi) + B(\phi)}}\,.
\end{equation}

One can show that the Hawking temperature is
\begin{equation}
\label{eq:tempgub5}
T = \frac{e^{A_h + B_h} |V'_h|} {4\pi}.
\end{equation}
and this can be shown to be \cite{Gubser:2008ny}
\begin{equation}
\label{eq:tempgub4}
T = \frac{\phi_h^{1/(\Delta -4 )}}{\pi R} \frac{V(\phi_h)}{V(0)} \exp{\left\{\int_0^{\phi_h} \left[ G(\phi) - \frac{1}{(\Delta-4) \phi} + \frac{1}{6 G(\phi)} \right] \right\}},
\end{equation}
where we used that $V(\phi \to 0) \to -12/R^2$, to leading order in $\phi$. Moreover, one can also find
\begin{equation}
\label{eq:Bconst}
B_h = \log \left[ \frac{4V(\phi_h)}{V(0)V'(\phi_h) L}\right] + \int_0^{\phi_h} \frac{d\phi}{6G(\phi)}\,.
\end{equation}

Let us continue with the thermodynamics. As Eq.\ \eqref{eq:gubseraction} is just the Einstein-Hilbert action coupled with some matter fields, the entropy density of the black brane is given by the area of the horizon 
\begin{equation}
\label{eq:entropydensitygub}
s = \frac{2\pi}{k_5^2} e^{3 A(\phi_h)} \,.
\end{equation}
Therefore Eqs.\ \eqref{eq:tempgub4} and \eqref{eq:entropydensitygub} give a thermodynamical equation of state  parametrized by $\phi_h$: $(T(\phi_h), s(\phi_h))$.  In particular, one can write the equation of state in terms the speed of sound
\begin{equation}
\label{eq:cs2gub}
c_s^2 = \frac{d \log T}{d \log s} = \frac{{d \log T}/{d\phi_h}}{{d \log s}/{d\phi_h}}\,.
\end{equation}

\subsection{Choice of the scalar potential}

In this framework, the potential $V(\phi)$ is chosen to match the QCD plasma thermodynamics at zero chemical potential. As mentioned above, the main restrictions on $V(\phi)$ are that near the boundary $\phi \to 0$, $V(\phi) \sim -12/R^2 + m^2 \phi^2/2$ while near the black brane horizon, $V(\phi) \sim V_0 e^{\gamma \phi}$. A simple, fairly featureless, potential that satisfies both conditions is
\begin{equation}
\label{eq:potmodel}
V(\phi) = -\frac{12}{R^2} (1+a\phi^4)^{1/4} \cosh (\gamma \phi) + b_2 \phi^2 + b_4 \phi^4 + b_6 \phi^6,
\end{equation}
where $\gamma$, $b_2$, $b_4$ and $b_6$ are the free parameters of the potential\footnote{Ref.\ \cite{Gubser:2000nd} obtained an important constraint that must be obeyed in order to avoid naked singularities that cannot be covered by a black brane horizon at finite temperature: $V(0) \geq V(\phi)$ for $\phi \neq 0$. For the choices of scalar potentials used here, within the range in temperature we were interested in, we did not find any naked singularities that could not be covered by a horizon.}. 

The parameter $a$ controls the nature of the thermodynamical phase transition; as we shall see, $a=1$ implies that the bulk theory has a Hawking-Page transition and thus the dual gauge theory has a first order phase transition - this class of models can be used to mimic the properties of the deconfinement transition in $SU(N_c)$ Yang-Mills theory \cite{Boyd:1996bx,Panero:2009tv}. On the other hand, $a=0$ implies that the dual gauge theory has a crossover phase transition and the model can be used to describe the thermodynamics of QCD with (2+1) light quark flavors \cite{Borsanyi:2010cj}. The models with $a=1$ and $a=0$ will be called here B1 models and B2 models, respectively.

The near-UV ($\phi \to 0$) mass $m^2$ of the bulk effective action can be extracted from Eq. \eqref{eq:potmodel}
\begin{equation}
\label{eq:effmass}
m^2 = -\frac{12\gamma^2}{R^2} + 2 b_2\,.
\end{equation}
On the other hand, as in the UV Eq.\ \eqref{eq:delta} holds, one obtains that $b_2$, $\Delta$, and $\gamma$ are not independent
\begin{equation}
\label{eq:deltarel}
b = \frac{6 \gamma^2}{R^2} + \frac{\Delta (\Delta - 4)}{2R^2}\,.
\end{equation}
In Table \ref{tab:param} we show the parameters for both models we consider in this work. We remark that in both models $\Delta = 3$, as used before in \cite{Ficnar:2010rn,Ficnar:2011yj,Ficnar:2012yu}. These two sets of parameters were chosen in order to fit lattice data for pure $SU(3)$ Yang-Mills theory and QCD, respectively - we shall display the numerical results for the thermodynamics in the corresponding sections for each model.

\begin{table}
\begin{center}
\begin{tabular}{ c || c c c c c c }
 \hline \hline                        
 & a & $\gamma$ & $b_2$ & $b_4$ & $b_6$ & $\Delta$ \\
 \hline
Model B1 & 1 & $\sqrt{2/3}$ & 5.5 & 0.3957 & 0.0135 & 3.0 \\
Model B2 & 0 & 0.606 & 0.703 & -0.12 & 0.0044 & 3.0 \\
  \hline \hline  
\end{tabular}
\caption{Parameters for the B1 (first order phase transition) and B2 (crossover phase transition) models. The last column shows the corresponding scaling dimension $\Delta$ of each model.\label{tab:param}}
\end{center}
\end{table}

\section{Debye screening mass and Polyakov loop in the B1 model}
\label{sec:B1model}

Let us start by the B1 model which possesses a first order deconfining phase transition and models the thermodynamics of pure $SU(N_c)$ Yang-Mills theory.

\subsection{Thermodynamics}

\begin{figure}
\centering
  \includegraphics[width=.5\linewidth]{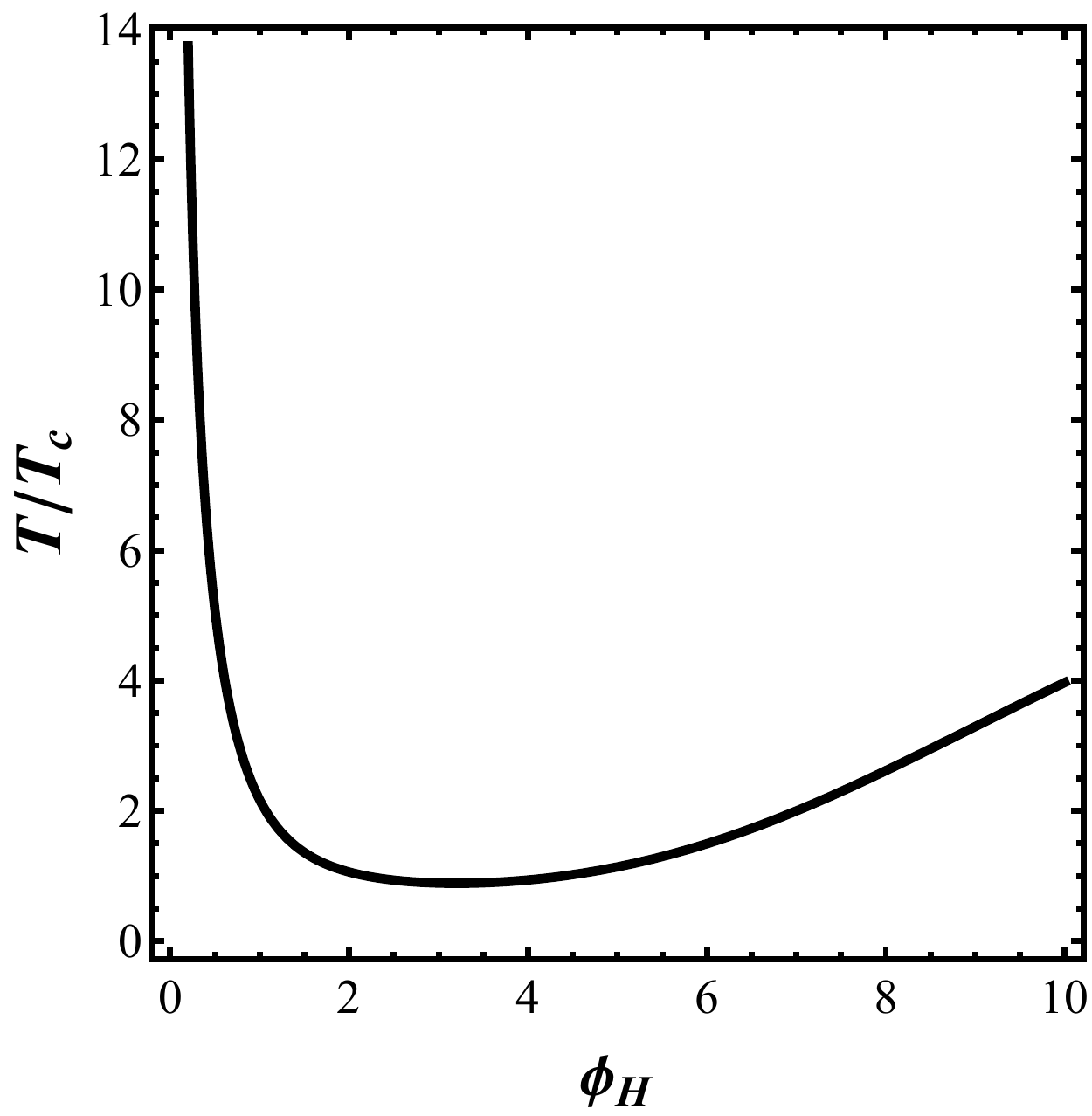}
  \caption{Temperature $T$ (normalized by the critical temperature $T_c$) as a function of the horizon position in the holographic coordinate $\phi_h$ for the B1 model.}
  \label{fig:TphiH-B1}      
\end{figure}

To obtain the thermodynamics of this model we use Eqs.\ \eqref{eq:tempgub4} and \eqref{eq:entropydensitygub}. We start by presenting, in Fig.\ \ref{fig:TphiH-B1}, the temperature $T$ (normalized by the critical temperature $T_c$ for the first order transition) as a function of $\phi_h$. As in Model A, we have two characteristic temperatures. First, we have a minimum temperature $T_{min}$ (given by the minimum of $T$ in Fig. \ref{fig:TphiH-B1}) below which the black hole solution does not exist and the dominating bulk geometry corresponds to a thermal gas. The second distinctive temperature is the critical temperature, $T_c$, at which the pressure of the black brane solution vanishes. For temperatures $T$ such that $T_{min} < T < T_c$, the thermal plasma is in a (superheated) metastable phase. For the parameters we used, given in Table \ref{tab:param}, $T_{min} = 0.89 T_c$, with $\phi_{h,min} = 3.20$ and $\phi_{h,c} = 2.20$.

\begin{figure}
\centering
  \includegraphics[width=.5\linewidth]{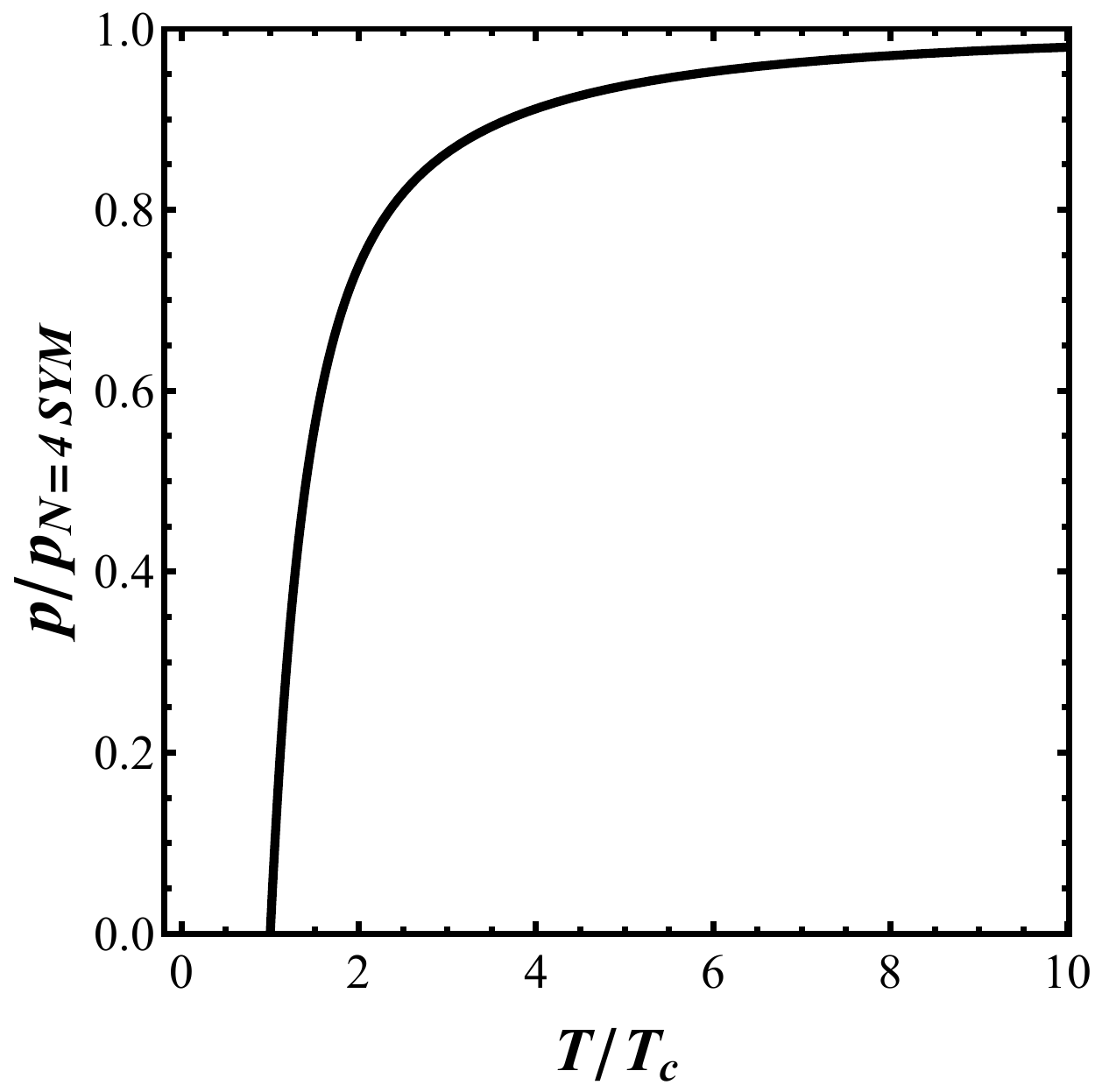}
  \caption[The pressure $p$ of the plasma for model B1.]{The pressure $p$ of the plasma for model B1, normalized by the $\mathcal{N} = 4$ SYM result, as a function of the normalized temperature $T/T_c$.}
  \label{fig:pressure-B1}      
\end{figure}

\begin{figure}
\centering
  \includegraphics[width=.5\linewidth]{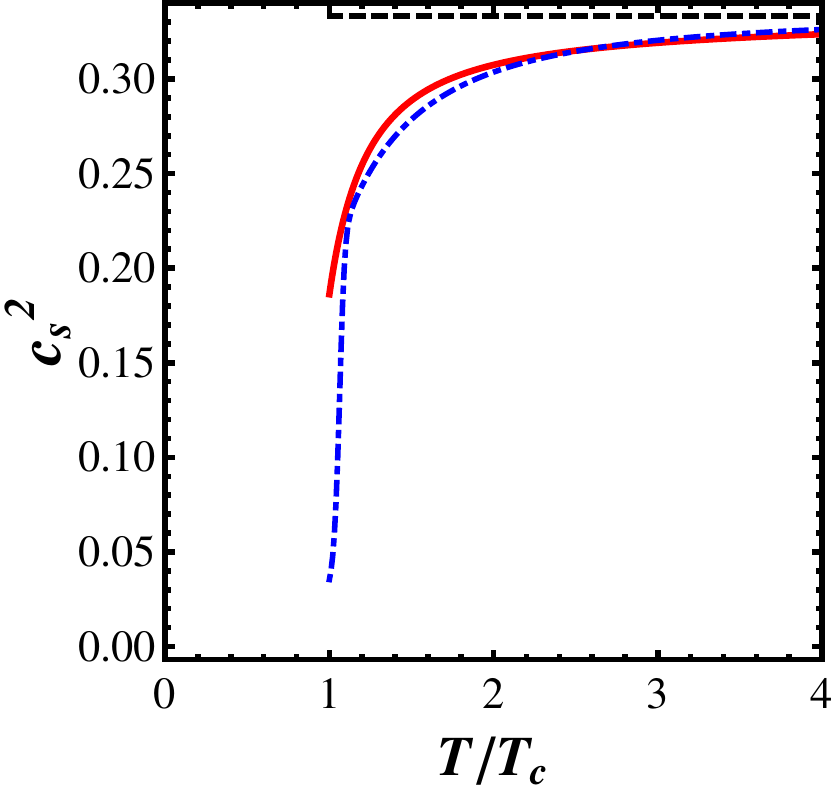}
  \caption[The sound speed squared of the plasma $c_s^2$ for model B1.]{(Color online) The speed of sound squared of the plasma $c_s^2$ for model B1 as a function of the normalized temperature $T/T_c$ (solid red curve), compared with $SU(3)$ Yang-Mills lattice results (dot-dashed blue curve) \cite{Boyd:1996bx}. The black dashed line is the CFT result, $c_s^2 = 1/3$.}
  \label{fig:cs2-B1}      
\end{figure}

From Eq.\ \eqref{eq:entropydensitygub} we evaluate the entropy density $s$ as a function of $\phi_h$. Using the results shown in Fig.\ \ref{fig:TphiH-B1}, one can eliminate $\phi_h$ and obtain $s$ as a function of $T$. With $s(T)$, one may proceed to evaluate all the thermodynamic functions. For instance, the pressure $p$ is given by
\begin{equation}
\label{eq:pressure-B1}
p = - \int_{\infty}^{\phi_h} s(x) T'(x) dx,
\end{equation}
while $c_s^2$ is given by Eq.\ \eqref{eq:cs2gub}. In Fig.\ \ref{fig:pressure-B1} we show the pressure $p$ (normalized by the $\mathcal{N} = 4$ SYM result) as a function of $T/T_c$. In Fig.\ \ref{fig:cs2-B1}, we compare the model results for the equation of state written in terms of $c_s^2$ with the corresponding lattice results for pure $SU(3)$ Yang-Mills \cite{Boyd:1996bx}. We see that the B1 model is in fair agreement with $SU(3)$ thermodynamics representing a quantitative improvement with respect to model A.

\subsection{Polyakov loop}
\label{sec:polyB1}

\begin{figure}
\centering
  \includegraphics[width=.5\linewidth]{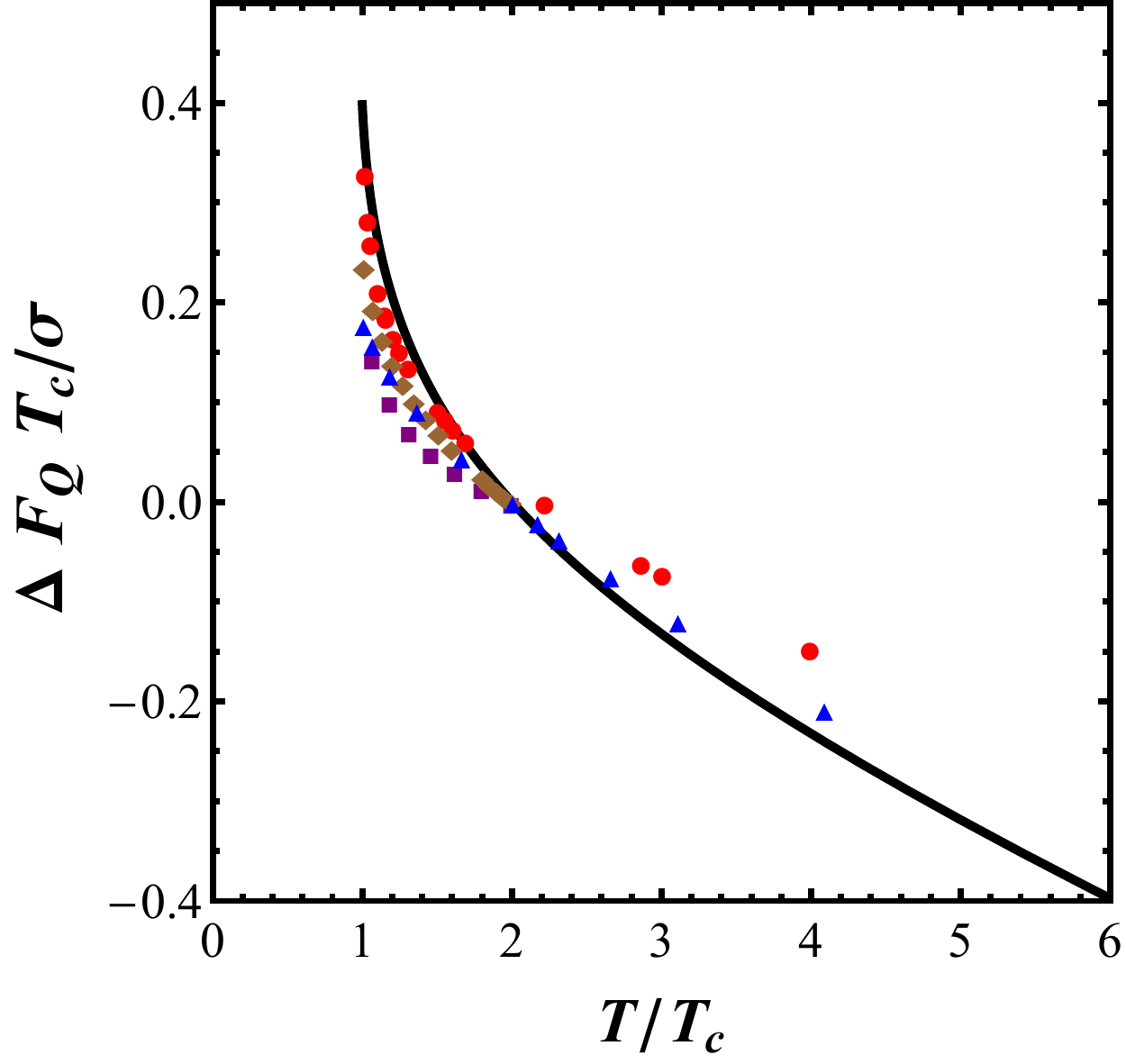}
  \caption{(Color online) $\Delta F_Q T_c/ \sigma = (F_Q(T)-F_Q(2T_c))T_c/\sigma$ as a function of $T/T_c$ for the model A (solid black line), model B1 (blue triangles), and for $SU(N_c)$ Yang-Mills \cite{Mykkanen:2012ri} with $N_c =$ 3 (red circles), 4 (purple squares), and 5 (brown diamonds).}
  \label{fig:FQ-B1}      
\end{figure}

The computation of the expectation value of the Polyakov loop proceeds as in \ref{sec:polyloop} using Eq.\ \eqref{eq:FQreg}.  This equation assumes that the geometry is in the conformal gauge; however, our numerical solution is obtained in the $\phi=z$ gauge. Thus, we need to perform a coordinate system change - the details of this gauge change can be found in Appendix \ref{sec:appendix-gauge}. Also, our geometry is given in the Einstein frame; to evaluate the Polyakov loop we have use the string frame. As in Model A, we assume that our geometry is related to some 5 dimensional subcritical string theory and the string frame metric is related to the Einstein frame metric by $g^s_{\mu \nu} = \lambda^{4/3} g_{\mu \nu}$, where $\lambda = e^{\phi}$.  A final remark is that in this model $b (y) \neq b_{(0)} (y)$ so that the cancelation that took place in Model A does not happen in this case. The regularized expression for the heavy quark free energy is 
\begin{equation}
\label{eq:FQregBclass}
\frac{F^{reg}_Q T_c}{\sigma} = \frac{T_c}{\Lambda b_s^2(y_{min})} \int^{y^c_h}_0 dy \, \left( b^2_s(y)-b^2_{(0),s}(y) \right)  - \frac{T_c}{\Lambda b_s^2 (y_{min})} \int^{y^c_h}_{y_h} dy\, b_s^2 (y),
\end{equation}
where $y = \phi$ in order to mantain the same notation used in \eqref{eq:FQreg}. In Fig.\ \ref{fig:FQ-B1} we show our numerical results for $\Delta F_Q \equiv F_Q(T) - F_Q(2T_c)$, comparing with lattice results for $SU(N_c)$ \cite{Mykkanen:2012ri}. One can see that model B1 follows more closely the lattice data in comparison that found for Model A.\footnote{{It should be noted that our models are built to study phenomena near the confinement/deconfinement transition from $T \sim T_c = 150 \mathrm{MeV}$ to $T \sim 3-4 \, T_c \sim 450-600 \, \mathrm{MeV}$. By construction, these models are strongly coupled in the UV. A reflection of this fact is that one cannot describe adequately both the Polyakov loop and the thermodynamics simultaneously at high temperatures, near the conformal regime, as argued in Ref. \cite{Zuo:2014iza}.}} 

\subsection{Debye screening mass}
\label{sec:mDB1}

We may now proceed to evaluate the Debye screening mass in the model B1. To obtain the Debye mass, we have to obtain the lowest eigenvalue $M^2$ of the corresponding Eq.\ \eqref{eq:axionsch}. As in the preceding subsection, this equation was written in the conformal gauge whereas our numerical solution for the metric is obtained in the Gubser gauge. The numerical procedure to find $m_D$ is exactly the same as described in \ref{debyesectionmodelA}. As in model A, we assume that the axion action is given by Eq.\ \eqref{eq:axionaction}, with the $\mathcal{Z}$ function given by the parametrization \eqref{eq:Zparam}. We use the same values of $c_4$ as in the study of model A, $c_4 = 0.1, 1$, and $10$.

\begin{figure}
\centering
  \includegraphics[width=.5\linewidth]{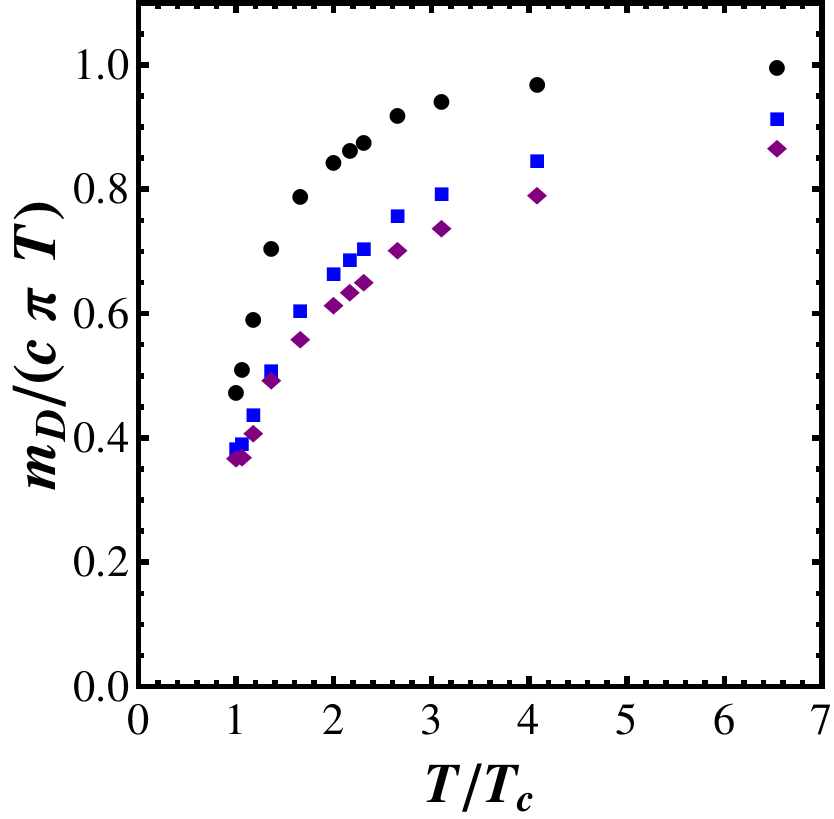}
  \caption[Debye screening mass for the B1 model.]{(Color online) Debye screening mass for the model B1, normalized by the $\mathcal{N} = 4$ SYM result $c\, \pi T$ (with $c = 3.4041$) as a function of $T/T_c$ for $c_4 = $ 0.1 (black circles), 1 (blue squares), and 10 (purple diamonds).}
  \label{fig:mD-B1}
\end{figure}

The numerical results for the Debye screening mass in this model are presented in Fig.\ \ref{fig:mD-B1}. As in the case of model A, $m_D/T$ has a discontinuity at $T=T_c$ where it jumps from 0 to a finite value $m_D/(c\pi T) \sim 0.35$ (somewhat higher than the jump in model A to $m_D/(c\pi T) \sim 0.2$). The value of the jump is not sensitive to the choice of $c_4$ and the overall behavior of $m_D/T$ as a function of $T$ saturates for large $c_4$. Note that we vary $c_4$ by two orders of magnitude and $m_D/T$ varies only by $\sim 20\%$ at high temperatures.

\section{Debye screening mass in the B2 model}
\label{sec:B2model}

\subsection{Thermodynamics}

\begin{figure}
\centering
  \includegraphics[width=.5\linewidth]{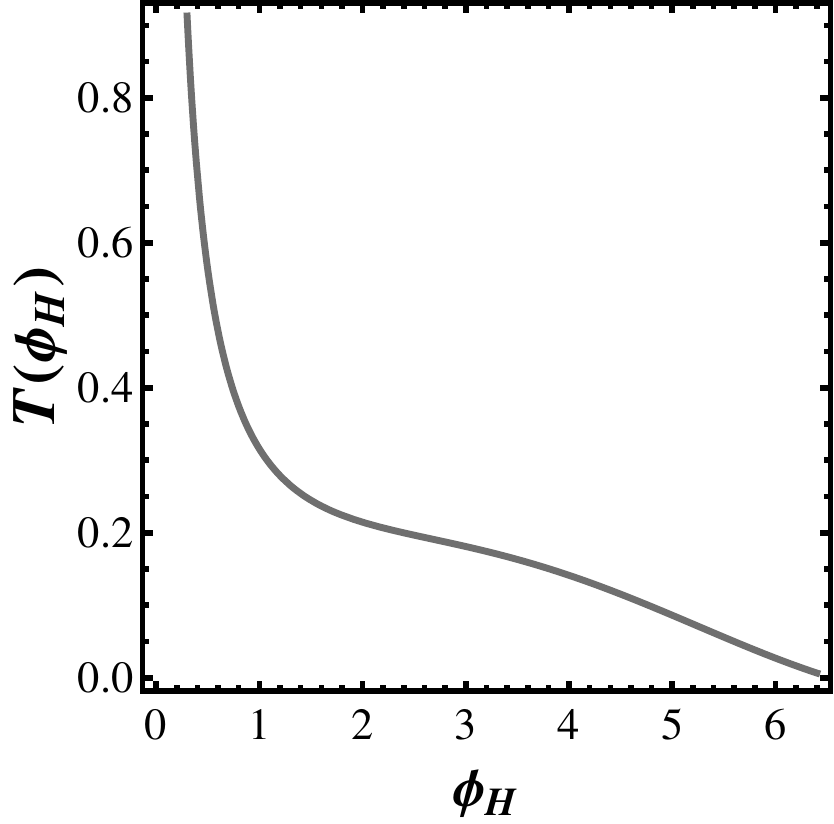}
  \caption{Temperature $T$ as a function of the horizon position in the holographic coordinate $\phi_h$ for the model B2.}
  \label{fig:TphiH-B2}      
\end{figure}

In this section we describe a choice of scalar potential that yields an equation of state for the holographic strongly coupled plasma that closely matches the lattice results for (2+1) QCD \cite{Borsanyi:2010cj}. The parameters for this potential can be found in Table \ref{tab:param}. For this model, the black brane solution always dominates over the thermal gas solution; thus, there is no metastable phase and no $T_{min}$. Also, there is no confinement at $T=0$. Moreover, the temperature $T$ as a function of $\phi_h$ is monotonically decreasing, as it can be seen in Fig.\ \ref{fig:TphiH-B2}. The pressure of the black brane phase is always positive and, thus, one cannot define a critical temperature $T_c$ as in Models A or B1. The phase transition in Model B2 is of crossover type; the thermodynamic quantities and their derivatives of all orders are continuous across the ``phase transition". In fact, the phase transition is characterized only by a sudden, but continuous, change of the thermodynamics properties. 

\begin{figure}
\centering
  \includegraphics[width=.5\linewidth]{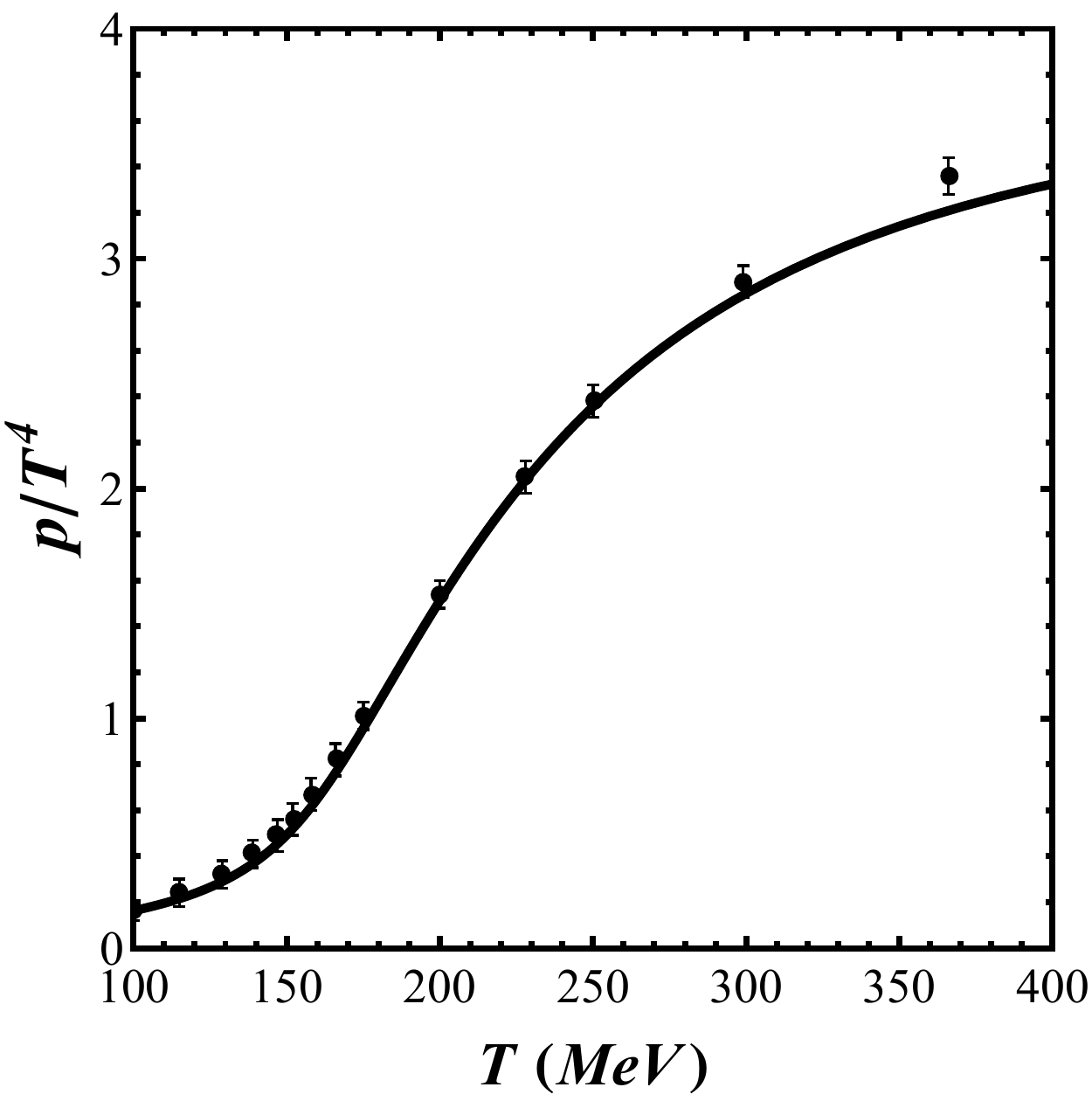}
  \caption{The pressure of the plasma $p/T^4$ as a function of the normalized temperature $T$, for the B2 model (solid curve), compared with (2+1) flavors $SU(3)$ QCD lattice results (data points) \cite{Borsanyi:2010cj}.}
  \label{fig:pressure-B2}      
\end{figure}

\begin{figure}
\centering
  \includegraphics[width=.5\linewidth]{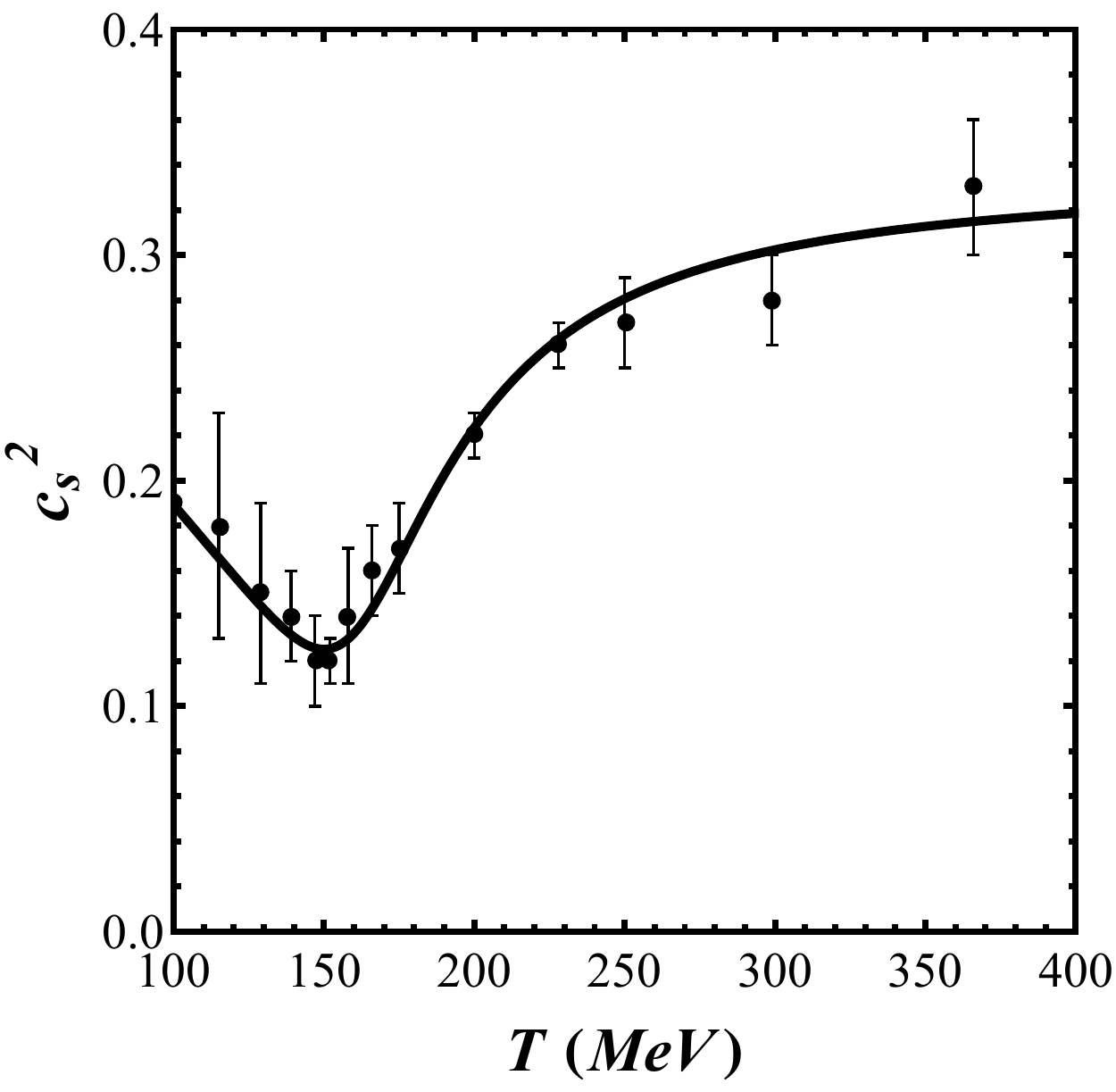}
  \caption{The speed of sound squared of the plasma $c_s^2$ as a function of temperature $T$, for the B2 model (solid curve), compared with (2+1) flavors $SU(3)$ QCD lattice results (data points) \cite{Borsanyi:2010cj}.}
  \label{fig:cs2-B2}      
\end{figure}

Model B2 gives a reasonable description of (2+1) QCD thermodynamics, as it can be seen in Fig.\ \ref{fig:pressure-B2} (pressure $p$ as a function of the temperature $T$) and in Fig.\ \ref{fig:cs2-B2} (equation of state in terms of $c_s^2$)\footnote{We use the position of the minimum of $c_s^2$ to set the scale of the temperature and express $T$ in MeV.}. The 5-dimensional Einstein's constant $G_5=0.501$ is chosen to reproduce lattice data for the pressure in Fig.\ \ref{fig:pressure-B2}. We also note that this model provides a quantitative description of the norm of the expectation value of the Polyakov loop found on the lattice \cite{Ficnar:2010rn}.

\subsection{Debye screening mass}

\begin{figure}
\centering
  \includegraphics[width=.5\linewidth]{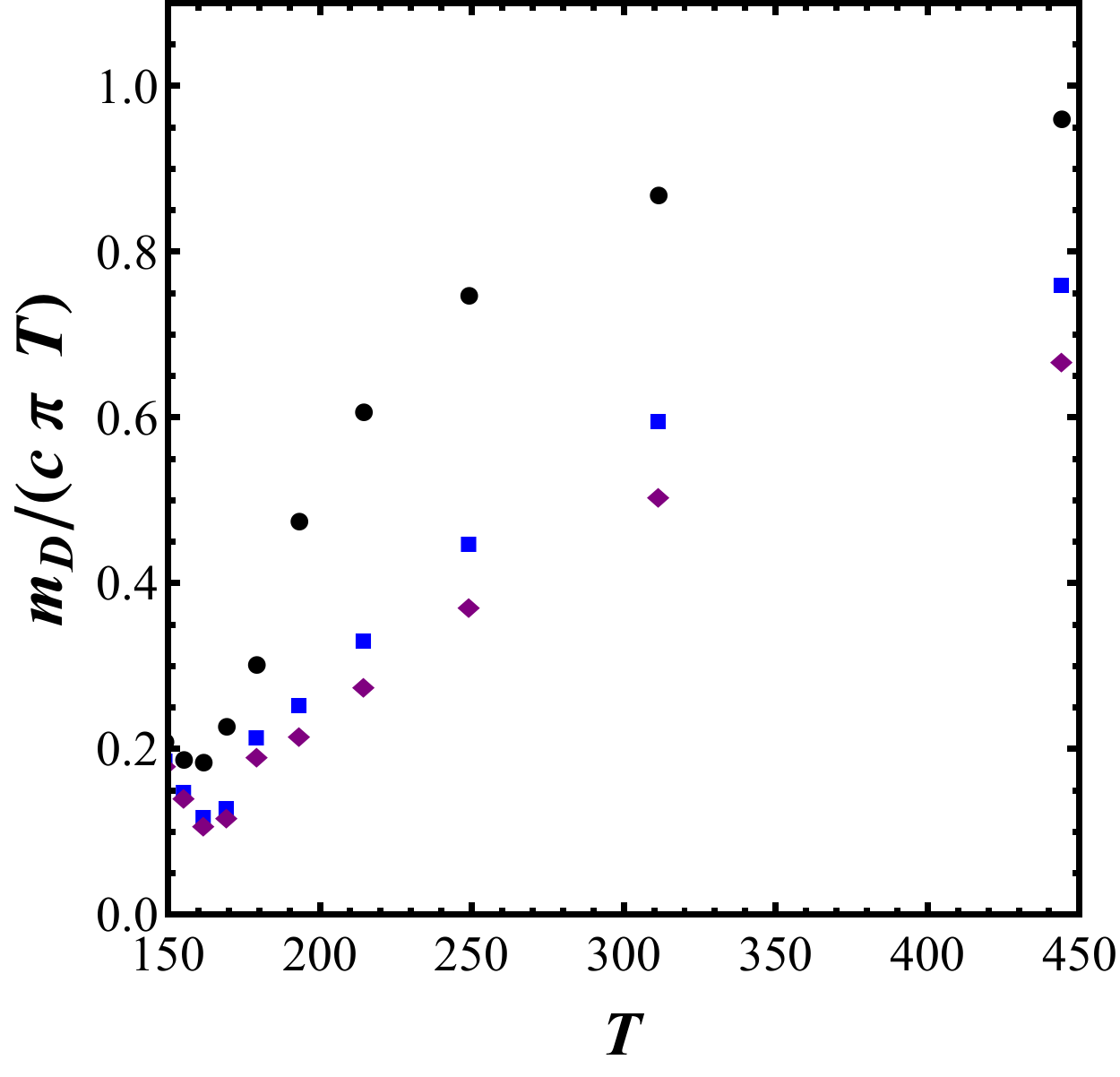}
  \caption{(Color online) Debye screening mass $m_D$ for the model B2 with a crossover transition, normalized by the $\mathcal{N} = 4$ SYM result $c\, \pi T$ (with $c = 3.4041$) as a function of the temperature $T$ for $c_4 = $ 0.1 (black circles), 1 (blue squares) and 10 (purple diamonds).}
  \label{fig:mD-B2}      
\end{figure}

Following the same procedure employed in previous sections, we may now evaluate the Debye screening mass as a function of the temperature in this model. The results are shown in Fig.\ \ref{fig:mD-B2}. The Debye screening mass $m_D/T$ has a local minimum around $T \sim 150$ MeV showing a similar temperature dependence found for $c_s^2$ (Fig.\ \ref{fig:TphiH-B2}). This minimum means, intuitively, that the plasma gets less screened (more transparent) to the strong interaction between colored heavy probes near the phase transition. Once again, larger values of $c_4$ show convergence and imply a faster rising to the conformal result (in this case by varying $c_4$ by two orders of magnitude the high $T$ values of $m_D/T$ vary by $\sim 30\%$).

\section{Debye mass dependence with $\eta/s$ - Gauss-Bonnet gravity}
\label{sec:gaussbonnet}

\subsection{Action and background Geometry}

As a final application of the holographic evaluation of the Debye screening mass, we consider a class of bulk actions that include curvature squared corrections to the supergravity action that violate the shear viscosity bound $\eta/s \geq 1/4\pi$ \cite{Kovtun:2004de}. The action for these gravity theories, also called Gauss-Bonnet gravity \cite{Zwiebach:1985uq}, is given by
\begin{align}
\label{eq:gaussbonnetaction}
S=&\frac{1}{16 \pi G_5} \int d^5x \sqrt{g} \left[ \left( \mathcal{R} + \frac{12}{R^2} \right) + \right. \nonumber \\ & \left. + \frac{\lambda_{GB}}{2} R^2 \left(\mathcal{R}^2-4 \mathcal{R}_{\mu \nu} \mathcal{R}^{\mu \nu} + \mathcal{R}_{\mu \nu \rho \sigma} \mathcal{R}^{\mu \nu \rho \sigma} \right) \right],
\end{align}
where $\mathcal{R}_{\mu \nu \rho \sigma}$ is the Riemann curvature tensor and $\lambda_{GB}$ is a constant. In Eq.\ \eqref{eq:gaussbonnetaction}, the first term is the usual second order Einstein-Hilbert action with the addition of the cosmological constant term. The constant $\lambda_{GB}$ is a measure of the size of the higher derivative corrections. The specific form the curvature squared corrections in \eqref{eq:gaussbonnetaction} implies that the metric fluctuations in a given background still follow second order equations. 

The action \eqref{eq:gaussbonnetaction} has an exact black brane solution \cite{Cai:2001dz}
\begin{equation}
\label{eq:metricgaussbonnet}
ds^2 = \frac{R^2}{z^2} \alpha^2 \left(f_{GB}(z) d\tau^2 + d\vec{x}^{\,2} +\frac{dz^2}{f_{GB}(z)} \right),
\end{equation}
where the scaling factor $\alpha$ is defined by
\begin{equation}
\label{eq:agaussbonnet}
\alpha^2 = \frac{1}{2} \left(1+ \sqrt{1-4\lambda_{GB} } \right)
\end{equation}
and the blackening factor $f_{GB}$ is given by
\begin{equation}
\label{eq:fGBgaussbonnet}
f_{GB} (z) = \frac{1}{2 \lambda_{GB}} \left[1-\sqrt{1 - 4\lambda_{GB} \left(1-\frac{z^4}{z_h^4} \right) } \right].
\end{equation}
We choose our coordinate system to write the background in a Poincar\'e patch-like form. The $z$ coordinate of the black brane horizon corresponds to the simple root of $f_{GB}(z)$, $z_h$. The temperature of the black brane solution is given by $T = \alpha /(\pi R^2 z_h)$. Comparing with Eq.\ \eqref{eq:metricgaussbonnet}, one can see that the scaling factor means that the AdS radius is now given by $\alpha R$. Finally, the 't Hooft coupling $\lambda$ in this case is $\lambda = \alpha^4 R^4/\alpha'^2$. The specific forms of $a$ and $f_{GB}$ imply that $\lambda_{GB} < 1/4$. Another constraint is given by imposing causality at the boundary, which implies $\lambda_{GB} \leq 9/100$ \cite{Brigante:2008gz}.

The shear viscosity/entropy density ratio $\eta/s$ in this model is related to $\lambda_{GB}$ by \cite{Brigante:2007nu}
\begin{equation}
\label{eq:etasgaussbonnet}
\frac{\eta}{s} = \frac{1}{4\pi} (1- 4 \lambda_{GB}).
\end{equation}
If $\lambda_{GB} > 0$ one has $\eta/s < 1/4\pi$ - the conjectured viscosity bound for gauge theories with gravity duals is then violated. Imposing $\lambda_{GB} \leq 9/100$ implies $\frac{4 \pi \eta}{s} \geq 16/25$.

\subsection{The Debye screening mass}

We have not specified the string theory construction that leads to Gauss-Bonnet gravity but such a discussion can be found in \cite{Buchel:2008vz}. The only field that can contribute to the channel used to define the Debye mass is the axion, which is once again trivial in this background. The action for the axion fluctuations \eqref{eq:axionaction} including only two derivatives is (this is still a conformal system and, thus, $\mathcal{Z}=1$)
\begin{equation}
\label{eq:axionactiongaussbonnet}
S = \frac{\alpha}{32 \pi G_5} \int d^5x \, e^{5\mathcal{A}} (\partial a)^2,
\end{equation}
where $\mathcal{A}(z) = \log (R/z)$. Apart from the constant factor of proportionality $\alpha$ in the action, this is the same action that would be obtained with a background of the form \eqref{eq:bgmetric}. So our equation of motion is still Eq.\ \eqref{eq:axionsch}, with $\mathcal{B} = 3/2 \log (z/R)$. As in Section \ref{sec:adsdebye} we use the dimensionless variable $y=z/z_h$, which yields the dimensionless mass $\tilde{M} = M/(\pi T)$.

Also, one can check that in this case the potential $\mathcal{V}(y)$ in Eq.\ \eqref{eq:axionsch} has the same asymptotic form near the boundary, namely $\mathcal{V}(y \to 0) = 15/(4y^2)$ - the leading term in $1/y$ is not changed. So, the asymptotic solutions are the same and all the tools used in \ref{sec:adsdebye} can be applied in this case without modifications. To obtain the Debye screening mass as a function of $\eta/s$, we analyze several values of $\lambda_{GB}$ and then use Eq.\ \eqref{eq:etasgaussbonnet} to obtain the corresponding values of $\eta/s$.

We also compare our numerical results with the phenomenological procedure pursued in Ref.\ \cite{Finazzo:2013rqy}. In that paper, we have evaluated in the strongly coupled plasma dual to Gauss-Bonnet gravity the  expectation value of the rectangular Wilson loop operator at finite temperature, which yields the potential energy $V_{Q\bar{Q}}$ of a heavy quark-antiquark pair that depends on $\eta/s$ \cite{Noronha:2009ia}. Using fits for the real part of the potential of the form
\begin{equation}
\label{eq:kmsmodelF-conf}
\frac{{\rm Re}\,V_{Q\bar{Q}}}{\sqrt{\lambda}T} = -\tilde{C}_1 \,\frac{e^{-\frac{\tilde{m}_D}{T} (LT)}}{(LT)^\delta}+\tilde{C}_2,
\end{equation}
where $L$ is the interquark distance while $\tilde{C}_1$, $\delta$, and $\tilde{m}_D$ were taken as fit parameters (we note that $\tilde{C}_2 = -1/\alpha^2$ by our regularization procedure) we found an estimate for the Debye screening mass $\tilde{m}_D$. For $\lambda_{GB} = 0$ we found $\tilde{m}_D = 3.79 \pi T$, in reasonable agreement with the result of Eq.\ \eqref{eq:n4debye}.

\begin{figure}
\centering
  \includegraphics[width=.5\linewidth]{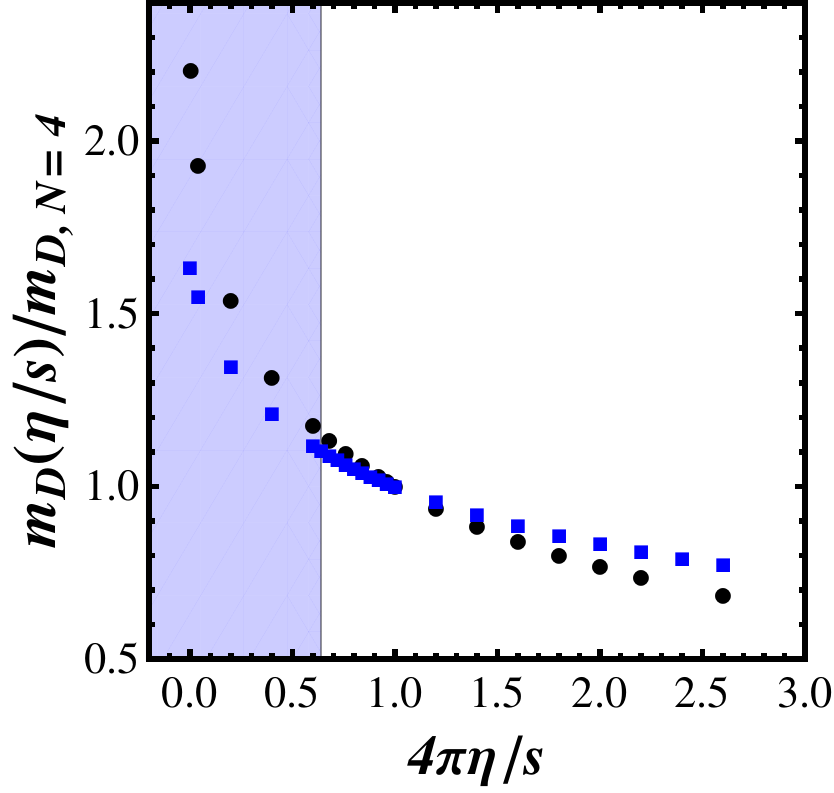}
  \caption[Debye mass $m_D$ for the Gauss-Bonnet gravity dual, as a function of $\eta/s$.]{(Color online) Debye mass $m_D$ for the Gauss-Bonnet gravity dual, as a function of $\eta/s$, normalized by the $\mathcal{N}=4$ SYM value. The black circles correspond to the results obtained by computing the lightest $CT$ odd mode; the blue squares are the results obtained by fits to the heavy quark-antiquark potential evaluated holographically \cite{Finazzo:2013rqy}. The shaded region corresponds to values of $\eta/s$ which violate the causality bound \cite{Brigante:2007nu,Brigante:2008gz}.}
  \label{fig:debyemassgaussbonnet}      
\end{figure}

We present the results for the Debye mass $m_D$ (normalized by the SYM value) as a function of the $\eta/s$ in Fig.\ \ref{fig:debyemassgaussbonnet}. We note that we have not restricted our calculations to the interval $\frac{4 \pi \eta}{s} \geq 16/25$ as required by causality but considered, for completeness, $\eta/s \geq 0$. One can see that for increasing $\eta/s$ the interaction between colored external probes in the plasma is less screened. This is reasonable, at least from the point of view of a weakly coupled plasma since $\eta/s$ is roughly proportional to the mean free path of momentum isotropization of the plasma and changing $\eta/s$ does not change the number of degrees of freedom of the system. Thus, less screening should correspond to a larger mean free path and, thus, to a larger $\eta/s$. We also note the unexpected coincidence between the results obtained by finding the lightest $CT$ odd mode and those obtained following the simple phenomenological procedure using the heavy quark potential described in the previous paragraph.

\section{Discussion and Conclusions}
\label{sec:conclusions}

In this paper we have identified the Debye screening mass $m_D$ in non-Abelian gauge theories at strong coupling with the lightest CT-odd mode in the spectrum (associated with the operator ${\rm Tr} F_{\mu\nu}\tilde{F}^{\mu\nu}$), following Ref.\ \cite{Arnold:1995bh,Bak:2007fk}. We used this prescription to holographically evaluate the Debye screening mass for a class of gravity duals involving the metric and a scalar field. Besides the conformal cases of $\mathcal{N} = 4$ SYM at strong coupling and the gauge theory dual to Gauss-Bonnet gravity (where the scalar field in the bulk vanishes), we investigated in detail an analytic bottom-up model with a first order confinement/deconfinement transition (Model A), and two bottom-up holographic models that describe the thermodynamics of QCD as seen on the lattice - Models B1 (pure glue, first order phase transition) and B2 (QCD, crossover transition).

The calculation of $m_D/T$ in both models for a pure Yang-Mills plasma with a first order phase transition at $T_c$, models A and B1, revealed some interesting features. Both models approach the conformal limit for $T \gg T_c$ and exhibit relatively little sensitivity to the axion coupling prefactor $\mathcal{Z}$. The most remarkable feature of both models is the discontinuity of $m_D/T$ at the critical temperature $T_c$ - $m_D$ jumps from 0 in the thermal gas phase ($T<T_c$) to a nonzero value at $T=T_c$. This behavior for $m_D/T$ in a pure SU($N_c$) Yang-Mills plasma is consistent with previous lattice studies \cite{Nakamura:2003pu}.

We also computed the expectation value of the Polyakov loop in these models finding an impressive agreement with lattice results \cite{Mykkanen:2012ri} even for $N_c = 3$. Moreover, even Model A, which does not provide an adequate quantitative description of SU(3) thermodynamics, yields a reasonable description for the Polyakov loop. This suggests that the Polyakov loop is largely insensitive to a variation in the number of colors $N_c$ in a pure glue plasma and that even $N_c = 3$ may be reasonably described by a large-$N_c$ expansion \cite{Mykkanen:2012ri}. Moreover, it would be interesting to identify more clearly what is the specific nonperturbative mechanism present in these holographic models that is responsible for this simultaneous description of lattice QCD thermodynamics and the expectation value of the polyakov loop.

Model B2 provides a reasonable description of the thermodynamics of (2+1) QCD\footnote{We should, however, emphasize that the gauge theory described by this gravity dual does not strictly possesses fermions in the fundamental representation. Those can be included using D-branes in the bulk geometry \cite{Karch:2002sh,Kruczenski:2003be}. See Ref.\ \cite{Erdmenger:2007cm} for a general review and \cite{Alho:2013hsa} for a study of the Veneziano limit in bottom-up constructions.}. The Debye screening mass, correspondingly, satisfies $m_D(T) > 0$ strictly and is always continuous. Near the crossover phase transition region at $T \sim  150 \, \mathrm{MeV}$, we see a minimum of $m_D/T$ (Fig.\ \ref{fig:mD-B2}). This minimum resembles, qualitatively, that found for the speed of sound squared $c_s^2(T)$, as shown in Fig.\ \ref{fig:cs2-B2}. For all the models, A, B1, and B2 the conformal regime is reached from below; that is, $m_D(T) < c \pi T$. The minimum of $m_D/T$ near the phase transition may have consequences for the energy loss of colored probes in the plasma \cite{Dumitru:2001xa}. Also, such a minimum implies that correlations in the medium are less screened, which effectively increases the range of interactions and this may be responsible for the (expected) small value of $\eta/s$ around $T\sim 150$ MeV \cite{Hirano:2005wx,Csernai:2006zz,NoronhaHostler:2008ju,NoronhaHostler:2012ug}. Equivalently, in this temperature range the expectation value of the Polyakov loop becomes small and, within the framework of the semi-QGP model \cite{Pisarski:2000eq,Hidaka:2009hs}, such a reduction may also lead to a suppression of $\eta/s$ \cite{Hidaka:2008dr,Hidaka:2009ma}.

The Debye screening mass of $\mathcal{N} = 4$ SYM at strong coupling, $m_D = 3.4041 \, \pi T$, extracted using the procedure of Ref.\ \cite{Arnold:1995bh}, yields a result that is remarkably close to the crude estimate used in Ref.\ \cite{Finazzo:2013rqy} where fits to the heavy quark-antiquark potential gave $m_D = 3.79 \pi T$. However, this coincidence should be interpreted with caution since, as discussed in Ref.\ \cite{Finazzo:2013rqy}, the heavy quark-antiquark potential in $\mathcal{N} = 4$ SYM at strong coupling is not exponentially screened (for small values of $LT$) as required to obtain the Debye screening mass from $V_{Q\bar{Q}}$.

By considering a gravity theory with higher order derivatives such that the gauge plasma does not satisfy $\eta/s = 1/(4\pi)$, namely Gauss-Bonnet gravity, we have evaluated the dependence of $m_D/T$ with $\eta/s$, as shown in Fig.\ \ref{fig:debyemassgaussbonnet}. We found that in this case less screening is seen as $\eta/s$ is increased. It would be interesting to check this result in other strongly coupled gauge theories. In particular, one could consider gravity duals that correspond to gauge theories in which $\eta/s < 1/(4\pi)$ still in the context of applications to the quark-gluon plasma. For example, axion-induced anisotropic deformations of $\mathcal{N} =4$ SYM \cite{Mateos:2011ix,Rebhan:2011vd} or strongly coupled $\mathcal{N}=4$ SYM subjected to an external magnetic field \cite{D'Hoker:2009mm,Critelli:2014kra}. However, the prescription of Ref.\ \cite{Arnold:1995bh} cannot be straightforwardly applied to these theories because they are not invariant by $CP$ - $P$ invariance is explicitly broken by the inclusion of the axion field in Ref.\ \cite{Mateos:2011ix} and by the presence of an external magnetic field in Ref.\ \cite{Critelli:2014kra}. 

\acknowledgments

We thank M.~Panero for making available to us the lattice results for the Polyakov loop from \cite{Mykkanen:2012ri} and for discussions about the renormalization scheme dependence of Polyakov loops. We also thank A.~Dumitru for very insightful comments on the manuscript and A.~Ficnar for discussions regarding the numerical solutions of Einstein's equations. The authors thank Funda\c c\~ao de Amparo \`a Pesquisa do Estado de S\~ao Paulo (FAPESP) and Conselho Nacional de Desenvolvimento Cient\'ifico e Tecnol\'ogico (CNPq) for support.

\appendix

\section{Gauge choices for model B1 and B2}
\label{sec:appendix-gauge}

As mentioned in the main text, for the models B1 and B2, the Gubser gauge \eqref{eq:gubseransatz} while adequate for studying the thermodynamics is not convenient for evaluating Polyakov and Wilson loops or finding the glueball spectrum (as done in Appendix \ref{sec:appendix-spectra}). For these purposes, it is convenient to change to the conformal gauge given by
\begin{equation}
\label{eq:confgauge}
ds^2 = e^{2\tilde{\mathcal{A}}(z)} \left(\tilde{h}(z) d\tau^2 + d\vec{x}^{\,2} + \frac{dz^2}{\tilde{h}(z)}\right).
\end{equation}
Comparing Eq. \eqref{eq:confgauge} with Eq. \eqref{eq:gubseransatz}, we see that the following relation must hold among the metric functions
\begin{equation}
\label{eq:gaugerel}
\frac{dz}{d\phi} = e^{B - A}.
\end{equation}
We require that the asymptotic $AdS_5$ is located at $z=0$ and that the horizon is at $z=z_h$. The solution of Eq. \eqref{eq:gaugerel} that satisfies these requirements is
\begin{equation}
\label{eq:gaugerel2}
z(\phi) = \int_0^{\phi}  d\tilde{\phi}\, e^{B(\tilde{\phi}) - A(\tilde{\phi})}
\end{equation}
We can invert (numerically) Eq. \eqref{eq:gaugerel2} to get $\phi(z)$. Then, the functions $\tilde{\mathcal{A}}(z)$ and $\tilde{h}(z)$ are given by $\tilde{\mathcal{A}}(z) = A(\phi(z))$ and $\tilde{h}(z) = h(\phi(z))$. 

\section{Glueball spectra in model B1}
\label{sec:appendix-spectra}

In this section we compute the glueball spectra for model B1, which displays confinement at $T=0$. The parameters used in the scalar potential in this model are given in Table \ref{tab:param}.

Let us briefly review the numerical procedure for finding the vacuum geometry and the glueball spectra. One first numerically integrates the equations of motion \eqref{eq:masterT0} subject to the boundary condition \eqref{eq:boundaryT0}; then, we search,  numerically, for the eigenvalues of the Schr\"odinger's equation \eqref{eq:axionsch2}, as described in the main text.  To find the spectra, we change the metric from the $z=\phi$ gauge \eqref{eq:gubseransatz} to the conformal gauge, as described in Appendix \ref{sec:appendix-gauge}. The potential for the Schr\"odinger's equation is given by Eq.\ \eqref{eq:potential}, where $\mathcal{B}$ depends on whether we are dealing with the scalar $J^{PC} = 0^{++}$ glueballs, tensor $J^{PC} = 2^{++}$ glueballs, or pseudo-scalar $J^{PC} = 0^{-+}$ glueballs \cite{Gursoy:2007er,Kiritsis:2006ua}
\begin{align}
\label{eq:Bglueball}
\mathrm{(scalar)} \, 0^{++} \quad \quad & \mathcal{B}(z) = \frac{3}{2} \mathcal{A}(z) + \frac{1}{2} \log X(z), \nonumber \\
\mathrm{(tensor)} \, 2^{++} \quad \quad & \mathcal{B}(z) = \frac{3}{2} \mathcal{A}(z), \\
\mathrm{(axial)} \, 0^{-+} \quad \quad & \mathcal{B}(z) = \frac{3}{2} \mathcal{A}(z) + \frac{1}{2} \log \mathcal{Z}(\lambda(z)). \nonumber
\end{align}
In Eq.\ \eqref{eq:Bglueball}, $X(z)$ is defined by
\begin{equation}
\label{eq:Xdef}
X(z) \equiv \frac{d\Phi/dz}{3\mathcal{A}(z)}, 
\end{equation}
where $\Phi = \sqrt{3/8} \phi(z)$ while $\lambda(z) = e^{\phi(z)}$, with $Z(\lambda)$ still given by Eq.\ \eqref{eq:Zparam}. For a comparison with lattice results, we normalize the spectrum by the fundamental $0^{++}$ glueball mass.

\begin{figure}
\centering
  \includegraphics[width=.5\linewidth]{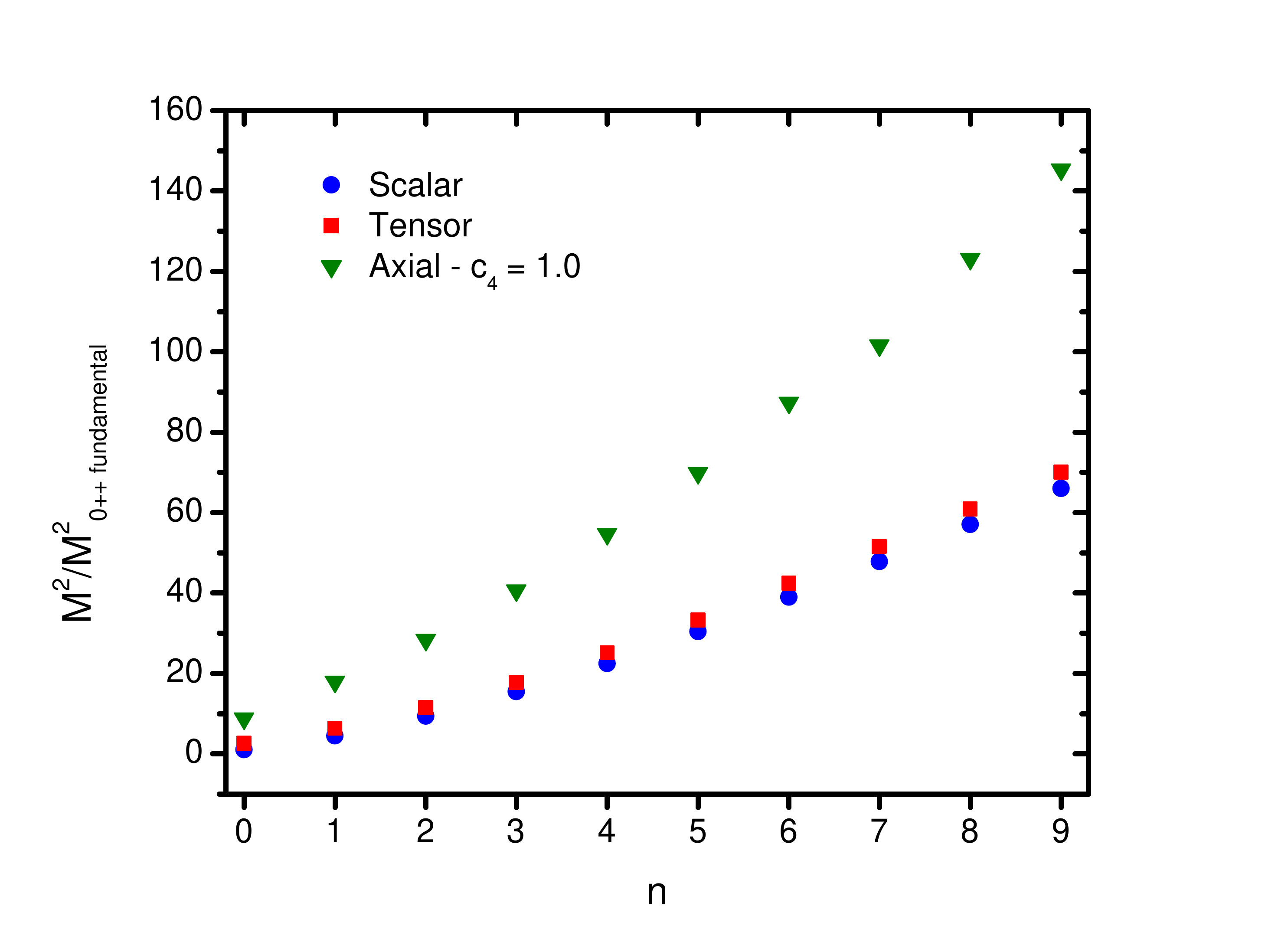}
  \caption[Glueball spectra in the B1 Model.]{(Color online) Glueball spectra in the Model B1. The glueball masses are normalized by the mass of the fundamental $J^{PC} = 0^{++}$ glueball. $n$ indicates the order of the excited state; $n=0$ is the fundamental state.}
  \label{fig:spectra-gubser}      
\end{figure}

\begin{figure}
\centering
  \includegraphics[width=.5\linewidth]{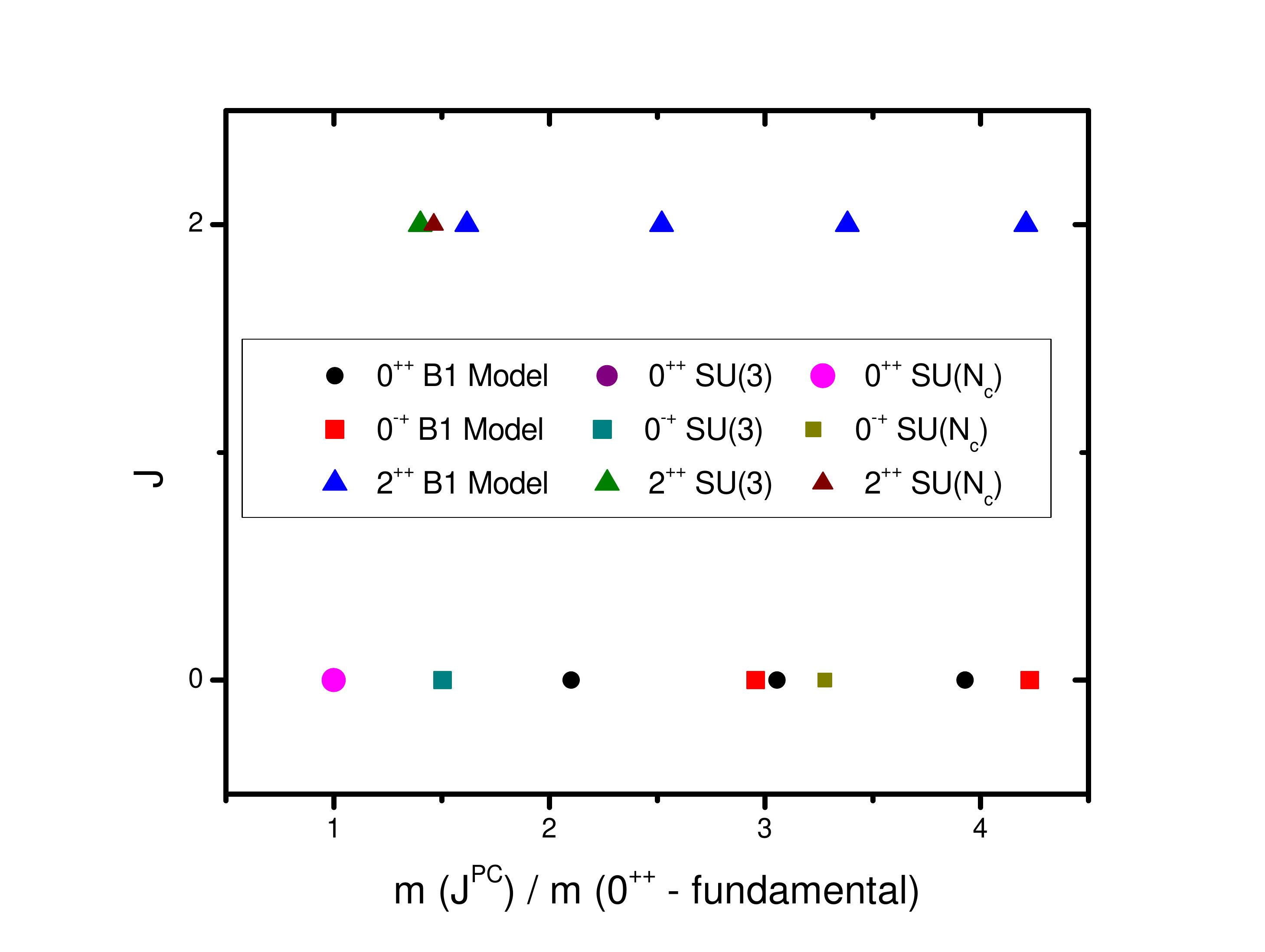}
  \caption{(Color online) Chew-Frautschi plot of the glueball spectra, comparing results from Model B1 with lattice results for SU(3) \cite{Morningstar:1999rf,Chen:2005mg} and large-$N_c$ $SU(N_c)$ \cite{Lucini:2001ej,Lucini:2010nv} Yang-Mills theory.}
  \label{fig:spectra-chewplot}      
\end{figure}

Our results are shown in Figs.\ \ref{fig:spectra-gubser} and \ref{fig:spectra-chewplot}. For comparison, we used lattice results for the glueball spectra in pure Yang-Mills with gauge groups $SU(3)$ \cite{Morningstar:1999rf,Chen:2005mg} and $SU(N_c)$ in the large-$N_c$ limit \cite{Lucini:2001ej,Lucini:2010nv}. We see in Fig.\ \ref{fig:spectra-gubser} that linear Regge trajectories are achieved for $n > 4$. Also, we note that the axial glueball has little sensitivity to the choice of $c_4$ - in the interval $c_4 = 0.1$ to $c_4 = 10$ the masses are almost degenerate. For this reason, in Fig.\ \ref{fig:spectra-gubser} we show only the results for $c_4 = 1$. Comparing with lattice results (Fig.\ \ref{fig:spectra-chewplot}), we see that reasonable agreement is found for the tensor glueball among all calculations. The axial glueball of the Model B1 and large-$N_c$ $SU(N_c)$ Yang-Mills are both reasonably close; however, both axial glueball masses are off by a factor of 2 when compared with the SU(3) Yang-Mills fundamental axial glueball. This contrasts with the results found for the holographic Polyakov loop in \ref{sec:polyB1}, where the results where relatively insensitive to $N_c$.

\end{document}